\documentclass[12pt]{article}
\usepackage{newtxtext,newtxmath}
\usepackage{graphicx}
\usepackage[a4paper,margin=2.5cm]{geometry}
\renewenvironment{abstract}
	{\quotation}
	{\endquotation}
\date{}

\makeatletter
\renewcommand{\fnum@figure}{\textbf{Figure \thefigure}}
\renewcommand{\fnum@table}{\textbf{Table \thetable}}
\makeatother
\usepackage{scicite}
\usepackage{url}
\usepackage{xcolor}
\usepackage[indent=1cm]{parskip}
\usepackage{pifont}

\def\scititle{Organic Electrochemical Transistor Arrays with Integrated Lipid-Sealed Femtolitre Chambers for Simultaneous Electrical and Optical Detection of Membrane Protein Activity}
\title{\bfseries \boldmath \scititle}
\author{
    S.~Kojima$^{1}$,
    S.~Rawat$^{1,2,3,4}$,
    M.~Sanchez~Miranda$^{1}$,
    J.G.~Gluschke$^{1}$,
    H.~Noji$^{5}$\and
    L.K.~Lee$^{2,3}$,
    A.P.~Micolich$^{1,\ast}$.\and
    \small$^{1}$School of Physics, University of New South Wales, Sydney NSW 2052, Australia.\and
    \small$^{2}$School of Biomedical Sciences, The University of New South Wales, Sydney NSW 2052, Australia.\and
    \small$^{3}$ARC CoE for Synthetic Biology, The University of New South Wales, Sydney NSW 2052, Australia.\and
    \small$^{4}$Australian Centre for Astrobiology, The University of New South Wales, Sydney NSW 2052, Australia.\and
    \small$^{5}$Department of Applied Chemistry, Graduate School of Engineering, The University of Tokyo, Tokyo, Japan.\and
    \small$^\ast$Corresponding author. Email: adam.micolich@nanoelectronics.physics.unsw.edu.au
}

\begin{document}

\maketitle

\begin{abstract} \bfseries \boldmath
We report a scalable method for fabricating arrays of fifty-two ion-sensitive PEDOT:PSS organic electrochemical transistors (OECTs) on glass coverslips, each incorporating an integrated fluoropolymer microwell sealed with a lipid bilayer for simultaneous electrical and fluorescence microscopy studies of membrane proteins. The architecture addresses a key limitation of existing OECT arrays by providing every transistor with an independent fluid reservoir sealed by its own isolated segment of lipid bilayer, thereby confining membrane protein diffusion to a defined sensing region. Microwells containing a dye-labelled $50$~mM KCl solution were sealed with lipid bilayers and overlaid with a dye-free $100$~mM KCl solution. Following addition of $\alpha$-hemolysin, formation of heptameric pores enabled K$^{+}$ ions to diffuse into the microwells and dye molecules to diffuse out, producing concurrent decreases in transistor conductance and fluorescence intensity. This integrated platform combines electrical and optical readout within independently addressable sensing sites and provides a versatile, scalable foundation for larger transistor arrays and the investigation of diverse membrane proteins reconstituted into lipid bilayers.
\end{abstract}

\section*{Introduction}
\noindent
The miniaturisation of semiconductor electronics to the nanoscale led naturally to work on interfacing electronic devices with biological entities ranging from whole cells, e.g., neurons, to the sub-cellular components vital to their function, e.g., ion channels in lipid bilayers~\cite{NoyAdvMat10, ZhangChemRev15, RivnaySciAdv17}. Early work at the sub-cellular level focussed on electrical detection of ion-channel~\cite{MisraPNAS09} and membrane protein~\cite{HuangNL10} activity. Efforts since have tended to focus in two directions. The first is the development of novel device platforms featuring relatively simple membrane models towards `hybrid' operations that merge electronic and biological functionality in creative new ways~\cite{SelbergCellSys18}. The second features more complex, biologically-realistic membrane models, often derived from living cells~\cite{LiuLangmuir20}, and is aimed at new electronic tools for {\it in vivo} and {\it in vitro} cell membrane studies for drug discovery and tissue engineering research applications~\cite{PitsalidisChemRev22}.

In parallel, the dual quests for improved ion-sensitivity and more biologically-favourable surfaces saw many researchers in bioelectronics shift away from inorganic semiconductor nanowire and graphene/carbon-nanotube field-effect transistor devices towards Organic Electro-Chemical Transistors (OECTs)~\cite{RivnayNatRevMater18}. The OECT features a conducting channel, typically consisting of a blend of the organic semiconductor poly(3,4‑ethylenedioxythiophene) and the dopant poly(styrene sulfonate). PEDOT:PSS is porous and hydrophilic providing an ideal surface for biological interfacing~\cite{RivnaySciAdv17} and the ability for ions from the electrolyte to diffuse into the transistor channel and act directly on the conductive PEDOT molecules, a process known as `volumetric' gating~\cite{RivnaySciAdv15}. The result is a transistor with a very high transconductance~\cite{KhodagholyNatComm13} and an ion-sensitivity well beyond the Nernst limit~\cite{GhittorelliNatComm18} because there is no gate oxide separating the electrolyte from the channel. The use of a PEDOT:PSS OECT as a sensor under a supported lipid bilayer was first demonstrated by Bernards {\it et al.}~\cite{BernardsAPL06}, and has become a commonly deployed configuration in the past 5-10 years for studying a range of membrane proteins in supported lipid bilayer systems~\cite{ZhangAFM16, LiuLangmuir20, PappaACSNano20, BaliACSAMI23, LuAdvSci24, McCoyAdvNanomedRes25, McCoyJMaterChemB25}.

A key feature of electronic devices interfaced to a supported lipid bilayer (SLB), be they PEDOT:PSS or alternatives such as Si nanowire transistors~\cite{MisraPNAS09, TunuguntlaAdvMat15, ChenNL19}, carbon-nanotube transistors~\cite{ZhouNatNano07, HuangNL10}, or Pd/PdH$_{x}$ electrodes~\cite{HemmatianNatComm16, Soto-RodriguezAdvMat16}, is that control over the lateral extent of the SLB is limited. While the SLB can, in principle, span the entire width of the well of fluid that constrains it, this is rarely the case. Instead, the SLB coverage area of a device depends heavily on the surface properties and the methods used in forming the SLB~\cite{TanakaNat05, HardyCOCIS13}. In cases where multiple devices or electrodes are placed under a single SLB, their response is never truly independent because they are all exposed to a common laterally-extended layer of fluid beneath the bilayer through which ions can readily diffuse. Additionally, membrane proteins also diffuse within the SLB such that over the course of a measurement, the exact ensemble of membrane proteins local to a given device/electrode evolves significantly. Given this, an obvious challenge in the scale-up of bilayer-based bioelectronic sensor arrays is to devise strategies to make the individual devices independent from a fluidic perspective. In other words, design the structure so that each device has its own integrated and independent reservoir of fluid that it senses from, sealed with its own isolated segment of lipid bilayer that constrains protein diffusion to a known limited range relative to the device. Previous work in this direction~\cite{TangACSNano21} has involved gluing multiple macroscopic glass fluid wells onto a single coverslip (e.g., see Fig.~1 of Tang {\it et al.}~\cite{TangACSNano21}), one per device or set of devices. A significantly more scalable alternative is to instead integrate the fluid well into the device itself using lithographic methods. This enables the fluid wells to be made orders of magnitude smaller, in our case with a diameter as small as $4~\mu$m. This integration step enables fabrication of arrays of independent microwell-integrated organic electrochemical transistors. By fitting within the field of view of a standard moderate-magnification epifluorescence objective, the arrays support simultaneous electrical and fluorescence microscopy while maintaining a compact, low-cost device footprint. Looking forward, our structures provide an exciting pathway to integrated high-density membrane-protein-augmented transistor arrays for potential applications as diverse as sensing, bioelectronics, drug discovery and neuromorphic computing.

\begin{figure}
\centering
{\includegraphics[width = 1.0\textwidth]{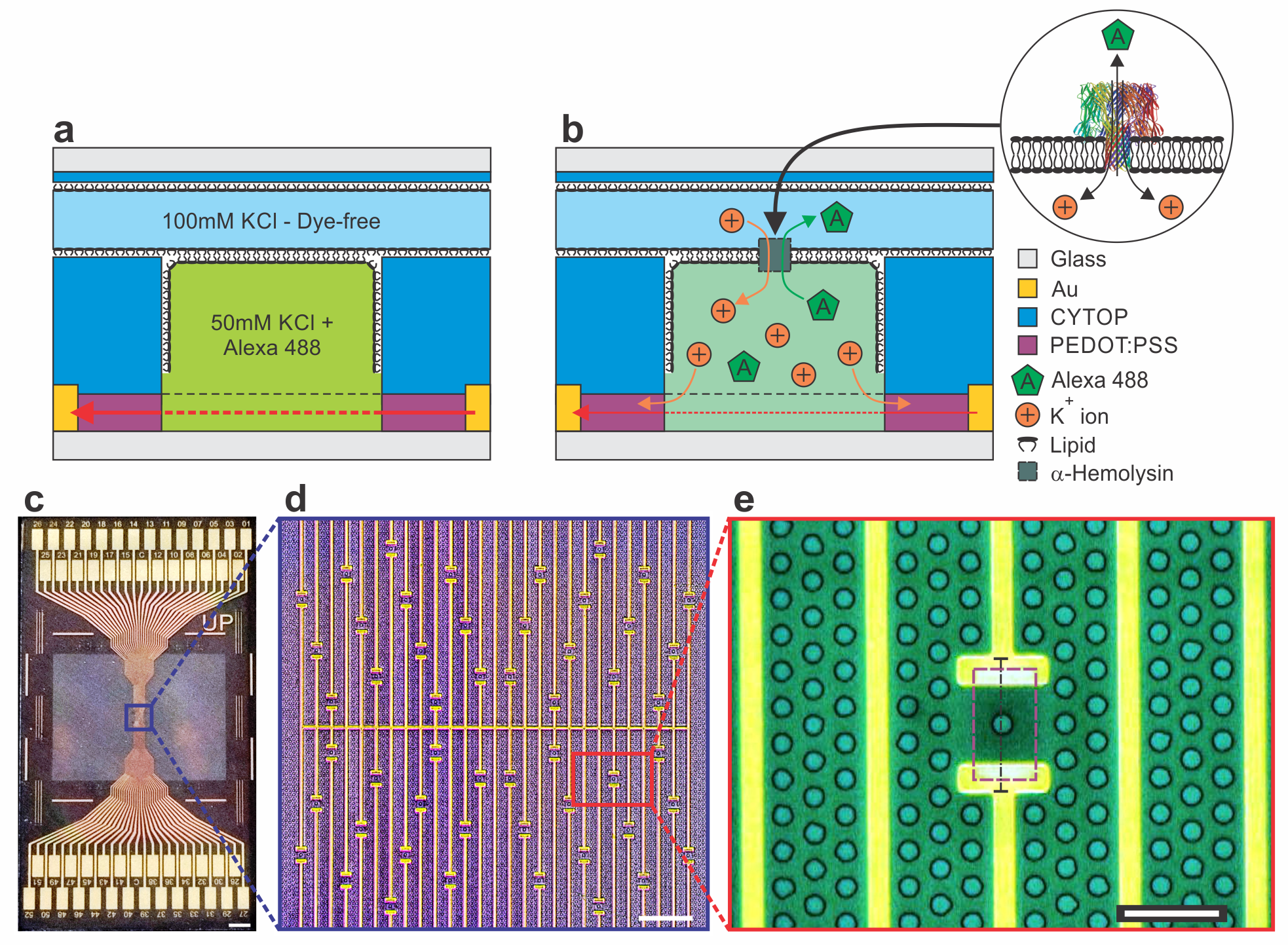}}
\caption{\textbf{Experiment Concept and Device Design.} ({\bf a,b}) Schematics of a single OECT device in the array ({\bf a}) before and ({\bf b}) after insertion of the $\alpha$-hemolysin pore, whereupon K$^{+}$ ions (orange) enter the well and Alexa-488 molecules (green) leave the well, reducing both the current (red arrow) through the PEDOT:PSS transistor channel and the microwell fluorescence intensity. ({\bf c, d, e}) Photographs of a completed $52$ OECT array slide at three different magnifications. The scale bars represent ({\bf c}) $2$~mm, ({\bf d}) $100~\mu$m, ({\bf e}) $20~\mu$m. The purple dashed rectangle in ({\bf e}) indicates the edges of the PEDOT:PSS channel. The black dot-dashed line in ({\bf e}) corresponds to the cross-section shown schematically in ({\bf a}) and ({\bf b}).}
\end{figure}

In this paper, we report a device containing an array of $52$ PEDOT:PSS OECTs, each of which is embedded at the bottom of a cylindrical microwell of depth $\sim~500$~nm and diameter $\sim~4~\mu$m. The microwells are patterned photolithographically and etched into a common layer of fluoropolymer that also electrically isolates the interconnects from the electrolyte. Under assay conditions, each microwell is sealed with its own isolated circular segment of lipid bilayer (see Fig.~1). Our concept builds on earlier work by Watanabe {\it et al.}~\cite{WatanabeNatComm14}, where a patterned layer of fluoropolymer on glass coverslip was used to generate large arrays of fluidically-independent femtolitre-capacity microwells, each sealed with an isolated segment of bilayer, for fluorescence-based sensing of membrane protein activity. The fluidically independent nature of these microwells arises from a unique bilayer morphology~\cite{WatanabeNatComm14, FujiwaraJChemPhys03, WangAnalyst12} generated by the fluoropolymer as shown in Fig.~1a -- the upper leaflet extends laterally across a large area of the coverslip (many microwells)~\cite{WatanabeSciRep14}, the lower leaflet for each microwell is a small segment that runs across the top of each microwell and down the microwell walls~\cite{WatanabePNAS18}. Our use of photolithographic patterning to define the microwells means that the $52$ microwells with integrated OECTs are contained within a larger array of over $3.2$~million non-device microwells in a hexagonally-close-packed array with centre-centre spacing of $\sim8~\mu$m. This abundance of microwells enables device studies at the few-to-single membrane protein per device limit using solutions at nanomolar concentration by effectively exploiting Poisson statistics~\cite{WatanabeNatComm14, ZhangAnalChem17}. The large array of non-device microwells is essential to the design because if those microwells are removed, i.e., we only have the $52$ microwells for the OECTs alone, populations at the few-to-single membrane proteins per microwell can only be obtained with impractically low solution concentrations, e.g., femtomolar or pg/mL, resulting in significant challenges with weighing and dilution precision in preparing the membrane protein assay solution. In this paper, we use the well-known membrane pore $\alpha$-hemolysin~\cite{AyubCOCB16, YingNatNano22} to demonstrate simultaneous electrical and fluorescence detection of protein activity using our devices via the process shown in Figs.~1a/b, however, other membrane proteins have been demonstrated via a similar approach, e.g., ATPase~\cite{WatanabeNatComm14}. We first use an aqueous-organic-aqueous exchange technique~\cite{WatanabeNatComm14} to seal each microwell with a buffer solution containing $50$~mM KCl and $20~\mu$M Alexa-488 dye on the inside (`inner' solution) and $100$~mM KCl on the outside (`outer' solution), as shown in Fig.~1a. We commence recording and then add $\alpha$-hemolysin monomers to the outer solution. The monomers insert into the lipid bilayer seals and spontaneously form heptameric transmembrane pores~\cite{BelmonteEBPJ87, WalkerJBiolChem92}. This results in the leakage of K$^{+}$ ions into the microwell to produce a change in electric current through the OECT, along with a slower leakage of Alexa-488 molecules out of the microwell to produce a fluorescence intensity change that serves as an effective experimental control for our study. Our work ultimately provides a path to simultaneous electrical and optical study of membrane proteins with ability to access the few-to-single protein limit with a statistically significant quantity of fluidically-independent ion-sensing OECTs per coverslip.

\begin{figure}
\centering
{\includegraphics[width = 0.6\textwidth]{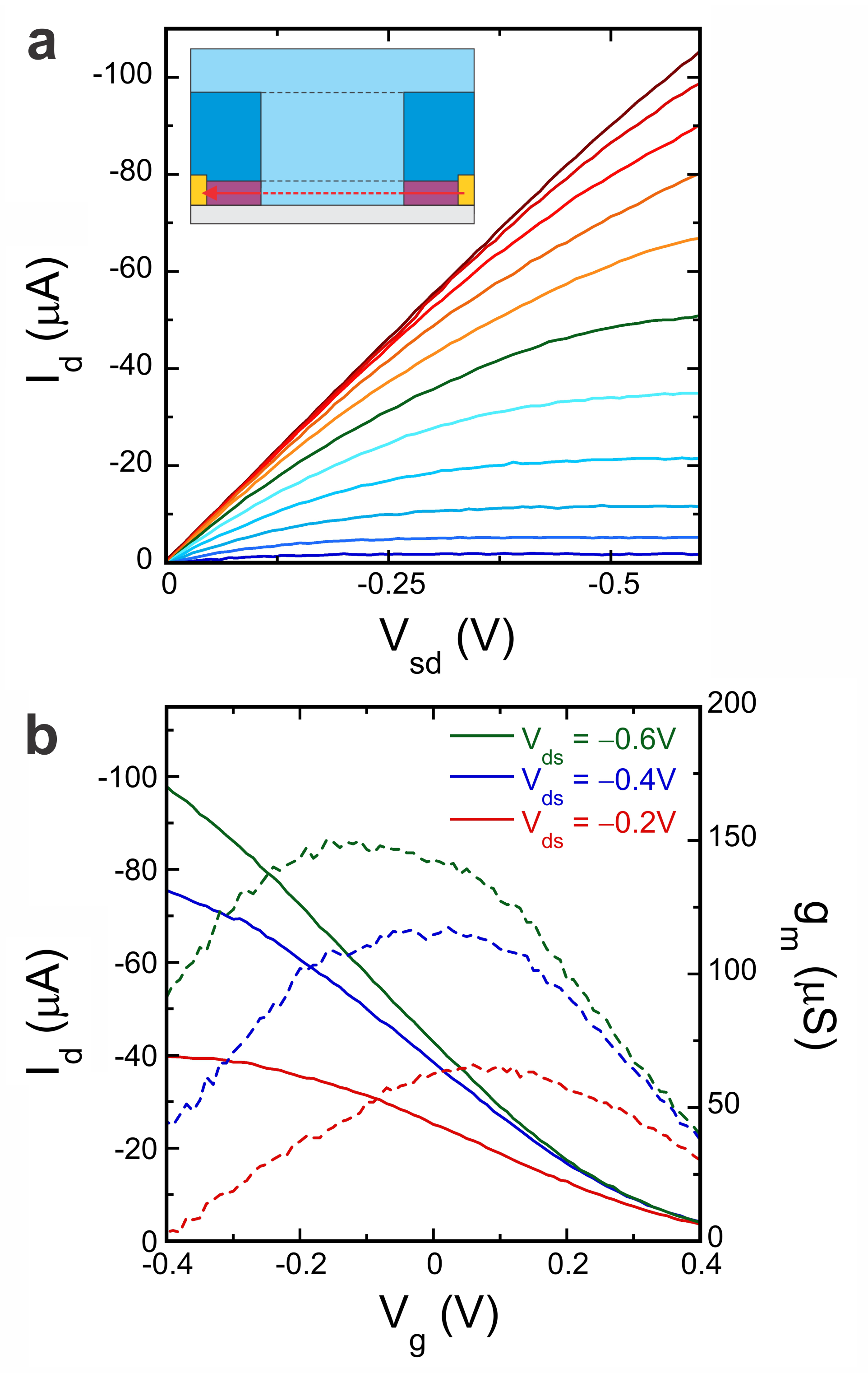}}
\caption{\textbf{Electrical Characterization of our OECTs.} ({\bf a}) Drain current $I_{d}$ vs source-drain bias $V_{sd}$ characteristics for gate voltages $V_{g}$ ranging from $-0.5$~V (top - brown) through $0.0$~V (green) to $+0.5$~V (bottom - purple) in steps of $0.1$~V for a typical OECT device. The inset shows a schematic of the device configuration for the measurements in Fig.~2 (and Fig.~3) where the microwell is unsealed and filled with KCl solution. ({\bf b}) Drain current $I_{d}$ (left-axis/solid lines) and transconductance $g_{m}$ (right-axis/dashed lines) vs gate voltage $V_{g}$ for three different source-drain biases $V_{sd}$. Data in both panels was obtained with [KCl] = $100$~mM and the gate voltage applied via an Ag/AgCl electrode.}
\end{figure}

\section*{Results}
\subsection*{Device Structure and Fabrication}
Figures~1c-e show photographs and micrographs of our device at three different scales. The devices are made on $22\times40$~mm rectangular $\#1.5$ glass coverslip using three stages of photolithography. In the first stage, we deposit a patterned layer of $15$~nm Al capped with $70$~nm Au, which serves as the source and drain contacts for the OECTs, the interconnects to the large pads for connection to instruments via spring-pins, and the fiducials for accurate alignment of the subsequent photolithography stages. In the second stage, we produce the $52$ PEDOT:PSS channels, which are aligned to the $18~\mu$m-wide source-drain gaps defined in the preceding stage. We spin-coat a mixture~\cite{StavrinidouAdvMat13} of $93.75\%$ PH1000-grade PEDOT:PSS, $5\%$ ethylene glycol, $0.25\%$ dodecylbenzene sulfonic acid (DBSA), and $1\%$ (3-glycidyloxypropyl)trimethoxysilane (GOPS) at $950$~rpm to achieve an approximately $100$~nm thick layer. This layer is patterned using negative photoresist to mask an O$_{2}$ plasma etch resulting in $26\times16~\mu$m channels with a contact footprint of $4\times16~\mu$m. The typical channel resistance of our as-fabricated transistors is $\sim 3-5$~k$\Omega$. In the third stage, we spin-coat a $\sim500$~nm thick fluoropolymer layer (AGC CYTOP CTL-809M), which serves the dual purpose of hosting the microwells and electrically insulating the metal layer from the buffer solution. The $4~\mu$m diameter microwells are arranged in a hexagonal-close-packed array with $8~\mu$m centre-centre spacing, filling the central $16\times12$~mm area of the coverslip (i.e., the area with a `cloudy' appearance at the centre of Fig.~1c). Any microwells that overlap metal features were manually removed from the array during mask design. We also remove any microwells that overlap the PEDOT:PSS channel, aside from the single microwell at the center of each PEDOT:PSS channel, as shown in Fig.~1e. The microwells are defined photolithographically, presenting a significant fabrication challenge because the strong hydrophobicity of the fluoropolymer counteracts photoresist adhesion. Watanabe {\it et al.}~\cite{WatanabeNatComm14} previously overcame this problem by using a highly viscous photoresist (AZ P4903), which adds significant cost, complex bake protocols to avoid resist cracking, and difficulties with edge-bead control, mask standoff, and resolution. We present here an alternate solution that enables standard viscosity photoresist to be used for patterning by exploiting the fact that CYTOP hydrophobicity can be manipulated by plasma chemistry~\cite{ChengSmall05}. Our complete recipe is given in the Materials and Methods~\cite{methods}, but briefly, we spin-coat and bake the CYTOP film which has a starting water contact angle of $108^{\circ}$ consistent with manufacturer specifications. We then briefly expose the CYTOP surface to O$_{2}$ plasma to make it more hydrophilic (contact angle $\sim95^{\circ}$), which enables adhesion of our standard photoresist (see Supplementary Fig.~S1). We then align, UV expose and develop that photoresist as usual prior to using it as the etch mask for a long O$_{2}$ reactive-ion etch (RIE) process in a parallel-plate configuration to define the microwells and expose the large contact pads at the two ends of the coverslip. This step is timed to etch to completion (i.e., expose bare hydrophilic glass in the unmasked regions), with the photoresist thickness tuned to ensure effective protection of the CYTOP surface in the masked regions. Finally, we strip the photoresist with acetone and expose the entire CYTOP surface to SF$_{6}$ plasma in the same parallel-plate RIE system for $5$~s to restore the CYTOP hydrophobicity. Our data in Supplementary Figs.~S1/2 show that hydrophobicity can be restored to well beyond that obtained immediately after spin-coating of the CYTOP; contact angles as high as $\sim140^{\circ}$ can be readily obtained. The restoration of strong hydrophobicity is vital to obtaining a high yield of bilayer-sealed microwells in our assay studies. Fabrication yields exceeding $98\%$ are readily obtained, with the main failure mode being imperfect metallisation at the corners of the coverslip due to poor resist planarisation (i.e., edge-beads) or broken interconnects due to contaminants (e.g., dust) in the first photolithography stage.

\subsection*{Electrical Characterization}
Figure~2a,b shows the typical drain current $I_{d}$ versus source-drain bias $V_{sd}$ and gate voltage $V_{g}$ characteristics for our OECTs (the associated gate electrode current $I_{g}$ data is shown in Supplementary Figure~S3). The microwells were not sealed with bilayer for these measurements. The OECTs were immersed in $100$~mM KCl solution using a custom PDMS well and the gate voltage $V_{g}$ applied via an Ag/AgCl microelectrode (InVivoMetric). Once the OECTs are immersed in KCl solution, we place the coverslip onto a metal block sitting in crushed ice for several minutes to draw the solution into the microwells and displace any trapped air, before returning the coverslip to room temperature for measurement (see Materials and Methods~\cite{methods} for further details on this technique). Our as-fabricated channel resistance is $3-5$~k$\Omega$ but tends to rise towards $>10$~k$\Omega$ after several gate sweeps because of the well-known tendency for performance degradation in PEDOT:PSS OECTs under repeated gate voltage cycling~\cite{KeeneMRSComm24}. This is why the channel resistance indicated by the characteristics in Fig.~2 is higher than that indicated by our ion-sensitivity and assay studies (Figs.~3 \& 5), where we avoid operating the device away from $V_{g} = 0$ unless strictly necessary. The typical distance between our OECTs is $50-100~\mu$m, depending on position in the array, and we observed no measurable crosstalk between devices during characterization. We obtain peak transconductances $g_{m} = \partial I_{d}/\partial V_{g} \sim 50-150~\mu$S for our devices. This corresponds to a geometry-normalised transconductance $g_{m,norm} = g_{m}/(Wd/L) \sim 8-25$~S/cm, where $W$, $L$ and $d$ are the channel width, channel length and PEDOT:PSS thickness, respectively. Our geometry-normalised transconductance is comparable to that normally found for PEDOT:PSS OECTs using a similar formulation (i.e., with EG, DBSA and GOPS additives)~\cite{InalNatComm17, KimNatComm18}. Although it is possible to improve the peak transconductance by optimising the geometry (i.e., balance of $W$, $L$, and $d$), we find $g_{m}$ to be sufficient for our purposes, and that the more important consideration is that the peak transconductance occurs as close to practicable to zero gate voltage. This enables us to keep our Ag/AgCl electrode at near-zero bias during our $\alpha$-hemolysin assay, avoiding both channel degradation and any issues that might arise from an applied electrochemical bias across the bilayer sealing the microwell, in particular, that the bias destabilises the bilayer causing it to rupture. For all studies that follow, our Ag/AgCl electrode is held at $V_{g} \sim 0$~V and simply acts to provide a ground reference for the buffer solution inside the flow-cell accordingly. Finally, we note that our OECTs show some gate hysteresis (see Supplementary Fig.~S4) as is common for PEDOT:PSS OECTs~\cite{StavrinidouAdvMat13, KumarAdvElecMat25}; we consider its impact on our experiment in later discussion.

\begin{figure}
\centering
{\includegraphics[width = 1.0\textwidth]{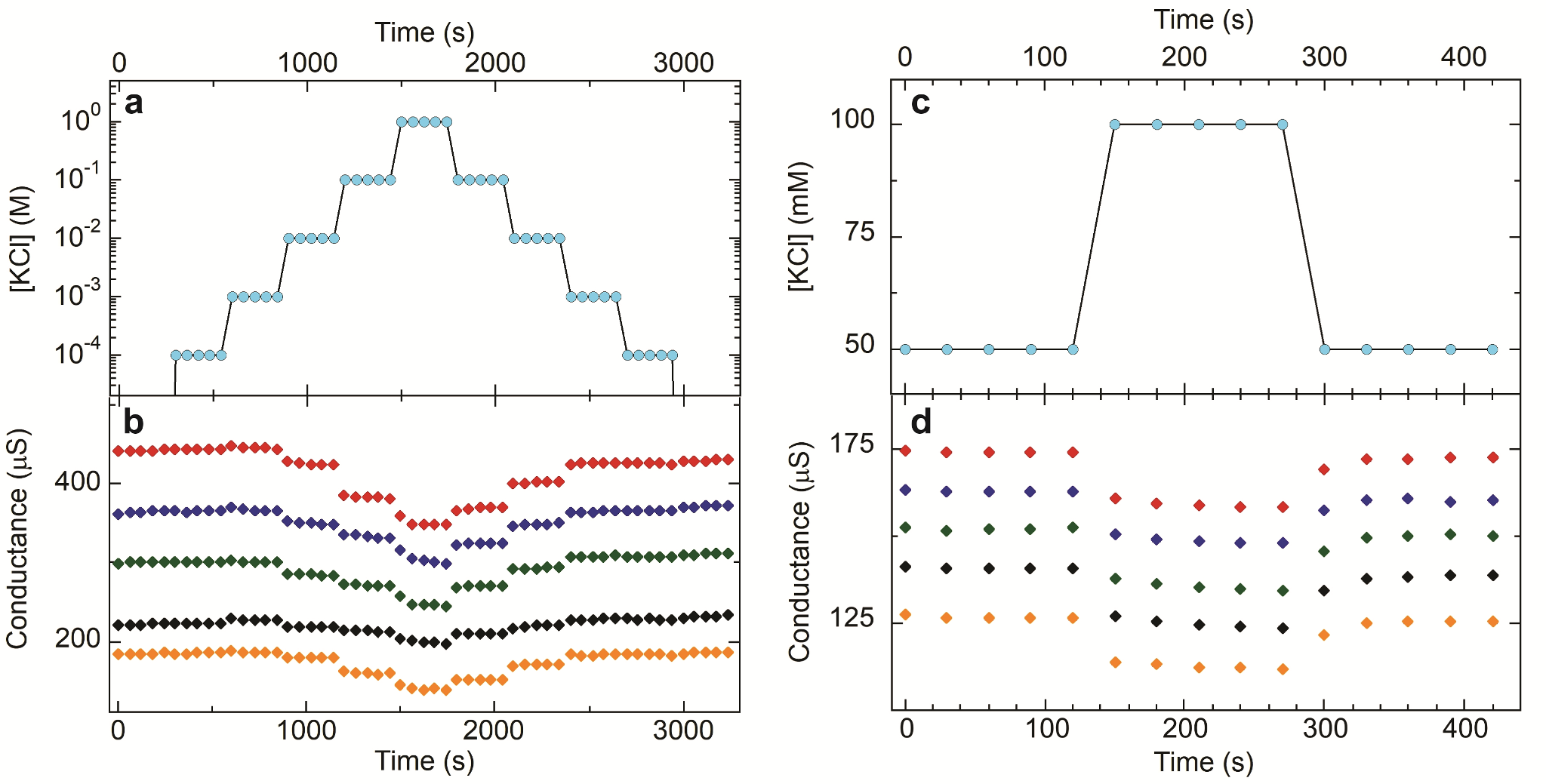}}
\caption{\textbf{Ion-Sensitivity Testing for our OECTs.} ({\bf a, b}) Plots of ({\bf a}) buffer KCl concentration and ({\bf b}) OECT channel conductance vs time for decadal progression of [KCl] for five typical OECTs on a single coverslip. Data offset for clarity: $\textcolor{red}\blacklozenge$ $\#1$, $0~\mu$S; $\textcolor{blue}\blacklozenge$ $\#7$, $-160~\mu$S; $\textcolor{green}\blacklozenge$ $\#31$, $-100~\mu$S; $\textcolor{black}\blacklozenge$ $\#47$, $-180~\mu$S; $\textcolor{orange}\blacklozenge$ $\#50$, $-200~\mu$S. ({\bf c, d}) Plots of ({\bf c}) buffer KCl concentration and ({\bf d}) OECT channel conductance vs time for the [KCl] change used in our $\alpha$-hemolysin assays for five typical OECTs on a single coverslip (different coverslip to ({\bf a, b})). Data offset for clarity: $\textcolor{red}\blacklozenge$ $\#16$, $0~\mu$S; $\textcolor{blue}\blacklozenge$ $\#22$, $-10~\mu$S; $\textcolor{green}\blacklozenge$ $\#31$, $-30~\mu$S; $\textcolor{black}\blacklozenge$ $\#34$, $-40~\mu$S; $\textcolor{orange}\blacklozenge$ $\#42$, $-50~\mu$S. Data in both panels was obtained with gate voltage held at $\sim0$~V via an Ag/AgCl electrode.}
\end{figure}

\subsection*{Ion sensitivity measurements}
Figures~3a/b shows the conductance response of five representative OECTs on a single coverslip of $52$ such OECTs to decadal step change in [KCl] spanning the range from $100~\mu$M to $1$~M. These studies are also performed without a bilayer sealing the microwells. The as-fabricated channel resistance can vary by several hundred ohms between the OECTs on a single coverslip. However, their response to ion concentration is sufficiently similar from device-to-device that a vertical offset in conductance is sufficient to obtain meaningful response comparison, as shown in Fig.~3b. We obtain an ion-sensitivity of order $12~\mu$S/dec over the $1$~mM -- $1$~M [K$^{+}$] range, with the ion-sensitivity diminishing for [K$^{+}$]$~<~1$~mM. This [K$^{+}$] operating range is more than sufficient for our experiments because we need to keep the [K$^{+}$] gradient across the bilayer to well below an order of magnitude to avoid problems with osmotic pressure, as discussed in the next section. Figures~3c/d show the typical device response to [K$^{+}$] changing between $100$~mM and $50$~mM, directly corresponding to the [K$^{+}$] of our inner and outer solutions for our assay studies (see Fig.~1a/b). We obtain a typical conductance change of $\sim 15~\mu$S between these two K$^{+}$ concentrations, much larger than the typical conductance uncertainty (i.e., noise threshold) of $0.2~\mu$S. This response is relatively consistent device-to-device despite the variations in as-fabricated channel resistance between devices on a given coverslip. We obtain yields of functioning and appropriately ion-sensitive OECTs exceeding $90\%$, with the few devices per coverslip that fail arising generally from cuts in the metal, either due to contaminants during the first photolithography stage or handling issues afterwards (e.g., accidental scratches during handling).

\subsection*{Effect of [K$^{+}$] gradient on the bilayer sealing the microwell}
The studies that follow were all performed after the microwells were sealed with bilayer. The process for sealing the microwells follows Watanabe {\it et al.}~\cite{WatanabeNatComm14} except that we implement it using a microfluidic system (Elveflow) rather than a motorised pipette (see Supplementary Fig.~S5). Briefly, a flow-cell is formed by using double-sided adhesive to seal a CYTOP-coated glass block with inlet/outlet ports over the coverslip. We introduce the inner solution into the flow-cell and then place the coverslip/flow-cell assembly onto a metal block sitting in crushed ice to draw the inner solution into the microwells and displace any trapped air. We flow a $1:1$ mixture of 1,2-dioleoyl-{\it sn}-glycero-3-phosphoethanolamine (DOPE) and 1,2-dioleoyl-{\it sn}-glycero-3-phosphoglycerol (DOPG) in chloroform at $4$~mg/mL total lipid concentration to form the inner leaflet of the bilayer in Fig.~4a. We then flow the outer solution into the flow-cell to form the outer leaflet of the bilayer in Fig.~4a, which results in each microwell being filled with inner solution and separated from the outer solution by the bilayer. Note that each microwell has an independent inner leaflet while the outer leaflet is common across the entire coverslip. This means each microwell/OECT is fluidically and ionically isolated from the others and any membrane protein that inserts into the bilayer sealing a given microwell remains geometrically constrained to that particular microwell as it diffuses in the bilayer~\cite{WatanabeNatComm14}. The coverslip/flow-cell assembly is removed from the cold metal block after the bilayer is formed and mounted on the microscope for optical/electrical studies. More complete details of the fluidics and microwell sealing process are given in the Materials and Methods~\cite{methods} and the Supplementary Information.

\begin{figure}
\centering
{\includegraphics[width = 1.0\textwidth]{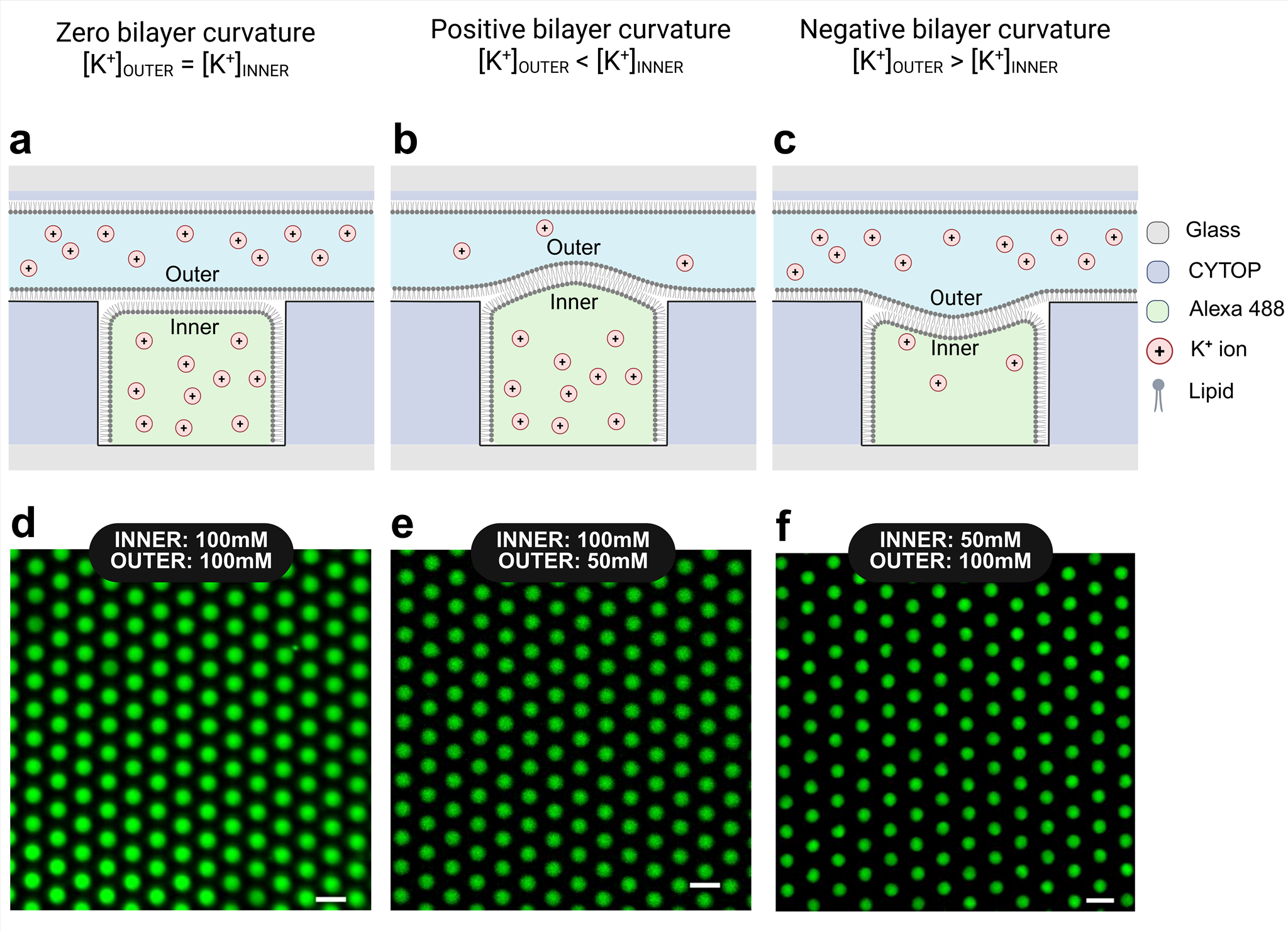}}
\caption{\textbf{Effect of K$^{+}$ Concentration on Bilayer Curvature.} ({\bf a}) Schematic and ({\bf b}) and fluorescence image of a bilayer-sealed microwell with equal [K$^{+}$] on both sides of the bilayer, which gives a flat bilayer at the centre of the microwell. ({\bf c}) Schematic and ({\bf d}) fluorescence images of a bilayer-sealed microwell with low [K$^{+}$] outside the microwell and high [K$^{+}$] inside the microwell, which gives a bilayer that bows outward (positive curvature). ({\bf e}) Schematic and ({\bf f}) fluorescence images of a bilayer-sealed microwell with high [K$^{+}$] outside the microwell and low [K$^{+}$] inside the microwell, which gives a bilayer that bows inward (negative curvature). All fluorescence images obtained with $20~\mu$M Alexa-488 in the inner solution and dye-free outer solution. The scale bar in {\bf d-f} represents $10~\mu$m.}
\end{figure}

We require a KCl concentration difference across the bilayer in order to generate an electrical signal in our experiment, and an important consideration is the effect this has on the inward/outward curvature of the bilayer across the top of the microwell due to the osmotic pressure difference (see Fig.~4b,c and Supplementary Fig.~S6a-c). From a purely device perspective, there is a strong incentive to maximise the [KCl] gradient across the bilayer because it would generate a more detectable signal in line with our data in Fig.~3b. However, the resulting [KCl] difference leads to a large osmotic pressure that either pushes the bilayer far outside the microwell ([KCl]$_{inner}~\gg~$[KCl]$_{outer}$) or collapses it inside along the wall and even floor of the microwell ([KCl]$_{outer}~\gg~$[KCl]$_{inner}$), as shown schematically in Supplementary Fig.~S6a-c. To demonstrate this, in Fig.~S6d-f we show fluorescence images for sealed microwells with a two order of magnitude [KCl] difference across them ($100$~mM/$1$~mM), which would be an ideal scenario from an electrical detection perspective. In Fig.~S6c/f, the bilayer sealing the microwells show pronounced outward bowing, which presents no major problems from a fluorescence intensity perspective, but makes these microwells highly susceptible to spontaneous rupture during subsequent fluidic flow steps and insertion of the $\alpha$-hemolysin. In contrast, the microwells in Fig.~S6a/b and d/e show that the bilayer has collapsed inward leaving the dye constrained to a thin rim around the bottom edge of the microwell. Thus, for our experiment to succeed, a careful trade-off is needed to ensure osmosis-driven curvature is minimal while an electrically detectable ion-concentration is still maintained. After extensive studies, we arrived at an optimum concentration difference of $100$~mM/$50$~mM, which gives a combination of reasonably constrained curvature (i.e., stable and equal microwell fluorescence intensity under static conditions), as shown in Fig.~4b/c and e/f, and detectable response in OECT conductance, as shown in Fig.~3d. It also keeps the ionic strength for the experiment in a biologically-relevant range. The only remaining decision for our assay studies was between negative and positive curvature. We chose negative curvature on the basis of studies by Fujii {\it et al.}\cite{FujiiACSChemBio15} showing that $\alpha$-hemolysin nanopore formation yield is relatively higher in membranes with negative curvature.

\subsection*{Simultaneous electrical and fluorescence recordings}
The remaining results and discussion focus on the main objective of the paper, which is the experiment shown schematically in Figs.~1a/b to simultaneously electrically and optically detect membrane protein activity. The experiment begins by following the same protocol for obtaining bilayer sealed microwells described previously using an inner solution with $50$~mM~KCl and $20~\mu$M Alexa-488 dye and a dye-free outer solution with $100$~mM~KCl. The coverslip is moved to a microscope and inspected to ensure that a majority of the device microwells, along with large areas of the non-device microwells surrounding them, are sealed and stable. There are three aspects we have found crucial to reliably obtaining reasonably high yield of stably-sealed microwells in our devices beyond the careful control of liquid flow noted by Watanabe {\it et al.}~\cite{WatanabeMMB18}. The first is to ensure good hydration of the PEDOT:PSS in our OECTs before the microwells are sealed otherwise liquid uptake by the PEDOT:PSS `pulls in' the bilayer either making measurements unstable or rupturing the seal entirely. To address this, we incubate our coverslips in inner solution for several hours before initiating the bilayer formation protocol. Second, the CYTOP surfaces need to be strongly hydrophobic including the CYTOP coating on the glass block forming the top of the flow-cell. This is achieved by performing a short SF$_{6}$ plasma etch before the assay and recoating the glass block intermittently as the coating degrades. Thirdly, fluid flow needs to be carefully managed, and in addition to being slow/continuous~\cite{WatanabeMMB18}, there can be no backflow or air bubbles entering the system.

\begin{figure}
\centering
{\includegraphics[width = 1.0\textwidth]{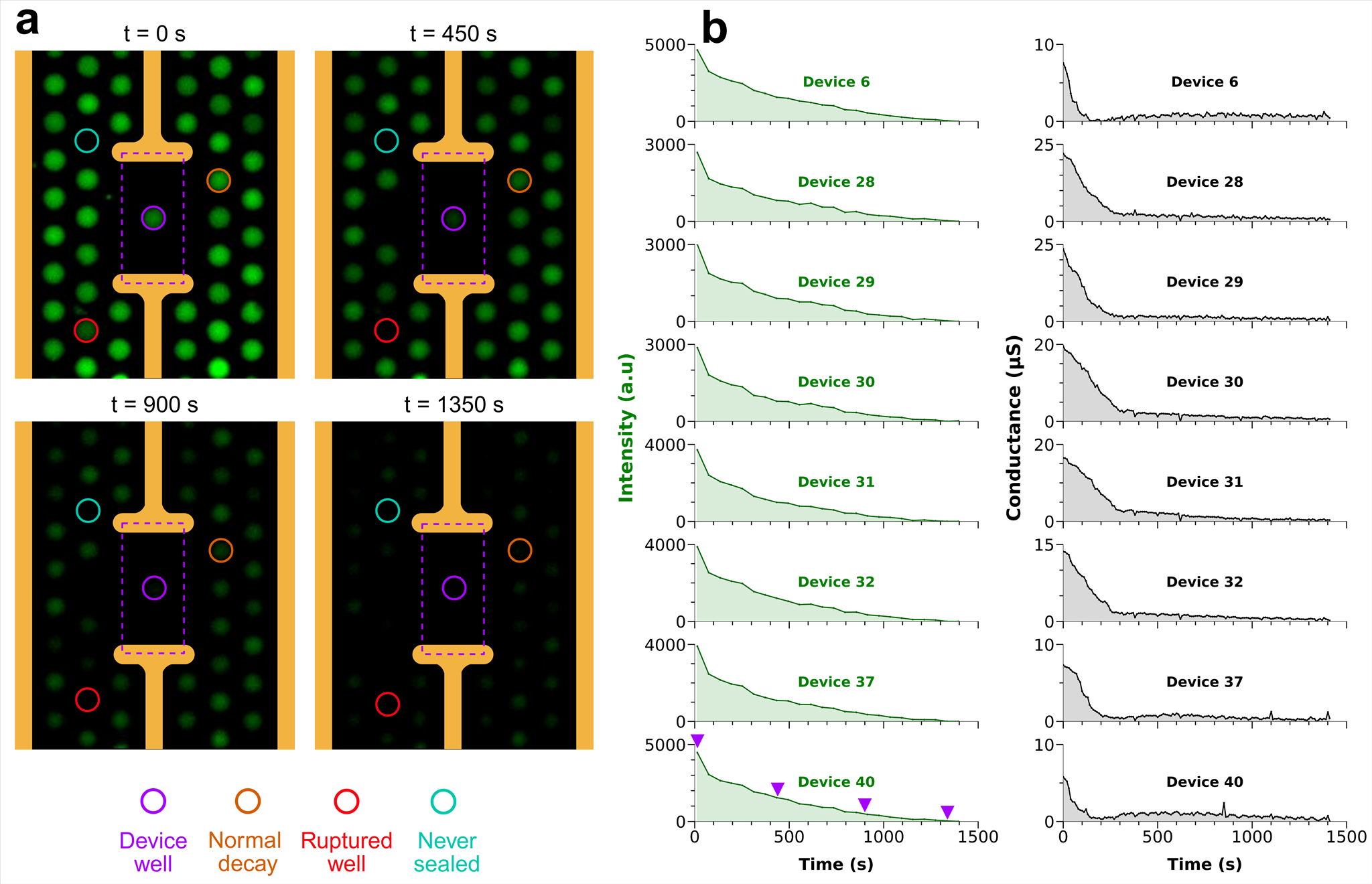}}
\caption{\textbf{Simultaneous Electrical and Optical Study of $\alpha$-hemolysin Activity.} ({\bf a}) fluorescence microscopy images of the region surrounding an OECT (Device 40 in ({\bf b})) at four times $t = 0$~s, $450$~s, $900$~s and $1350$~s during the assay. Four microwells are circled to indicate examples of: a device microwell with embedded $\alpha$-hemolysin (purple circle), a non-device microwell with embedded $\alpha$-hemolysin (brown circle), a non-device microwell with embedded $\alpha$-hemolysin that ruptures part way through the assay (red circle), and a non-device microwell that failed to seal with bilayer (cyan circle). The gold regions indicate where metal leads are and the underlying greyscale image has been converted to greenscale for clarity. The dashed purple lines indicate the edges of the PEDOT:PSS channel. ({\bf b}) Plots of fluorescence intensity (left column/green) and OECT conductance (right column/black) vs time $t$ for eight selected OECTs on a single coverslip. The intensity/conductance data has an offset of Device~6: $3993$~a.u./$237~\mu$S, Device~28: $2954$~a.u./$260~\mu$S, Device~29: $3382$~a.u./$259~\mu$S, Device~30: $2992$~a.u./$263~\mu$S, Device~31: $3374$~a.u./$231~\mu$S, Device~32: $3724$~a.u./$241~\mu$S, Device~37: $3675$~a.u./$236~\mu$S and Device~40: $4241$~a.u./$274~\mu$S subtracted for clarity. The four downward pointing purple triangles in the plot of fluorescence intensity vs time for Device~40 indicate the time-points for the four fluorescence microscopy images shown in ({\bf a}).}
\end{figure}

Once the microwells are sealed with bilayer, we flow outer solution containing $5~\mu$g/mL ($150$~nM) of $\alpha$-hemolysin monomer until the flow-cell is filled with this solution, whereupon we terminate the flow (this is our time $t=0$ in Fig.~5). The $\alpha$-hemolysin monomers insert into the bilayer where they assemble to form a heptameric pore~\cite{WalkerJBiolChem92, GouauxPNAS94} with diameter $\sim 2.6$~nm\cite{SongSci96}, sufficient for both K$^{+}$ ions and Alexa-488 dye molecules to pass through. This produces a simultaneous drop in OECT conductance and microwell fluorescence intensity, respectively. The large size difference between K$^{+}$ ions and Alexa-488 dye molecules means that the time-scales for their decay in concentration inside the microwell differ. Our diffusion calculations show that the half-time for Alexa-488 concentration decay is of order $30$~minutes, consistent with the actual timescale of fluorescence intensity assays in our work here and work by Watanabe {\it et al.}~\cite{WatanabeNatComm14}, while the expected timescale for [K$^{+}$] equilibration is of order a minute (see Supplementary Fig.~S7 and associated discussion). The much slower decay for the fluorescence signal in our experiment is fortuitous because it enables us to discriminate a rapid change in OECT conductance occurring by K$^{+}$ diffusion through the pore from the bilayer simply rupturing and rapidly releasing its contents into the outer solution.

In Figure~5a we show four fluorescence microscopy images of Device~40 from Fig.~5b at four different times $t$ after the flow-cell is filled with $\alpha$-hemolysin-containing buffer and pores start to form in the bilayers sealing the microwells. The device microwell is highlighted with a purple circle and shows a decaying fluorescence intensity as a function of time as Alexa-488 dye gradually leaks from the microwell via the $\alpha$-hemolysin pore. The majority of the microwells on a coverslip will show the same slow decay in fluorescence; one example for a non-device well is highlighted with a brown circle in Fig.~5a. There are three other possibilities for the behaviour of a given microwell. The first is that the bilayer seal on the microwell spontaneously ruptures at some $t > 0$, an example of such a microwell is highlighted with the red circle in Fig.~5a. These microwells show an instantaneous change in fluorescence intensity and OECT conductance that allow us to identify them and set them aside in the data analysis. The second are microwells that either failed to seal with bilayer or ruptured at some $t < 0$, an example is shown highlighted with the cyan circle in Fig.~5a. Such microwells start with low fluorescence intensity and OECT conductance and these remain low for all $t > 0$. The third is that the microwell does not have an $\alpha$-hemolysin pore form and the fluorescence intensity and OECT conductance start high and remain high for all $t > 0$. This case is rare in our measurement assays as we run with relatively high $\alpha$-hemolysin concentration to ensure as many of the device microwells host at least one $\alpha$-hemolysin, and therefore generate signal. However, an example is highlighted with a blue circle in the fluorescence images from a separate assay shown in Supplementary Fig.~S8b.

Figure~5b shows fluorescence intensity (green) and OECT conductance data (black) obtained simultaneously for eight different OECTs on a single coverslip. The four purple triangles above the fluorescence data for Device~40 indicate the time points corresponding to the four fluorescence images of Device~40 in Fig.~5a. The fluorescence intensity is obtained as the average pixel intensity for a circular region-of-interest with a diameter of 20~pixels centred on each microwell, in this case the device microwell for a given OECT. The electrical conductance is directly measured from the OECT under a source-drain bias of $-0.2$~V; for clarity in Fig.~5b, we have offset the OECT conductance by $\sim250~\mu$S (exact values given in caption). More complete details of both measurements are given in the Materials and Methods~\cite{methods} and the Supplementary information. Each device in Fig.~5b shows a fluorescence intensity decay that is approximately exponential consistent with diffusion of the Alexa-488 dye out of the microwell via the $\alpha$-hemolysin pore. This decay occurs over a period of order $20$~minutes, consistent with previous work by Watanabe {\it et al.}~\cite{WatanabeNatComm14}. The OECT conductance shows a sharp drop in the first $150$~seconds of the experiment consistent with an increase in the microwell [KCl] as K$^{+}$ ions enter the microwell via the $\alpha$-hemolysin (c.f. Fig.~3c/d). This shorter timescale for the conductance change is expected because the K$^{+}$ ion is much smaller than an Alexa-488 dye molecule but this effect not as drastic as our diffusion calculations in Supplementary Fig.~S7 would suggest alone -- a timescale difference of a factor of $\sim40$ is expected but only a factor of $\sim10$ is observed. We attribute this discrepancy partly to the fact that our diffusion calculations only treat diffusion through the $\alpha$-hemolysin into the microwell and not diffusion of K$^{+}$ ions from the microwell into the PEDOT:PSS channel. The PEDOT:PSS channel is buried under the CYTOP and only has a small interface area with the fluid inside the microwell, i.e., a cylindrical strip approximately $4~\mu$m in diameter and $\sim100$~nm high (see Fig.~1a). The remainder of the discrepancy comes from the fact that the diffusion of K$^{+}$ ions within the PEDOT:PSS channel will invariably differ from the fluid in the microwell. Although the mobility of K$^{+}$ ions in well-hydrated PEDOT:PSS can potentially be as high than the electrophoretic mobility in bulk water, the addition of a crosslinker, e.g., GOPS, can reduce this by well over an order of magnitude, and poor hydration can reduce it even further~\cite{StavrinidouAdvMat13}. An additional important aspect of our OECT response is that the K$^{+}$ ions need to diffuse laterally through several microns of a very thin film ($\sim100$~nm), ultimately to the edges of the OECT channel as shown in the schematic in Supplementary Fig.~S9, to produce the conductance change signal that we measure experimentally. The narrowings at the sides of the OECT channel produced by its `hole inside a rectangle' geometry are a key design feature that govern the rapid response of the OECT because these two small `side-segments' dominate the overall channel resistance. A careful optimisation of side-segment width relative to the microwell diameter could improve the future performance of these devices. The combination of OECT channel structure and K$^{+}$ ion diffusion in PEDOT:PSS also combine to govern the longer timescale `tail' behaviour of the conductance traces in Fig.~5b. One might expect the OECT conductance to remain fixed once the microwell [KCl] saturates at $100$~mM, but we often see the conductance continue to gradually reduce at a much slower rate, e.g., Devices~30, 31, and 32, or briefly rise again before commencing a gradual long slow decay, e.g., Devices~37 and 40. Although the side-segments dominate the conductance response, the overall volume of the channel subject to cation-driven doping/dedoping is much larger, meaning the overall time for the entire channel to arrive at ionic equilibrium and a truly static OECT conductivity can be much longer than the initial response. This behaviour is linked to the well-known gate hysteresis in OECTs~\cite{StavrinidouAdvMat13, KumarAdvElecMat25}, which we also observe in our devices (see Supplementary Fig.~S4). We find the temporal response in the electrical recording is strongly dependent on the pre-assay hydration state of the OECT channel, as highlighted in Supplementary Fig.~S9, where a device (blue trace) that was hydrated for only $3$~hours prior to use shows a long linear tail after the initial conductance response whereas another device (orange trace) hydrated for $5$~hours prior to use achieves a stable conductance after the initial conductance response. Note also that these two devices differ significantly in their actual channel conductivity despite being nominally identical from a fabrication perspective, indicating that the nanostructure of the PEDOT:PSS channel and its effect on K$^{+}$ ion diffusion likely also play an important role. This is not surprising, since as mentioned earlier, the only access to the OECT channel for hydration is a tiny cylindrical area of the microwell wall and hydration needs to occur out to PEDOT:PSS regions many microns away from this access. This suggests a pathway to improved OECT response in our usage, which is to reduce the source-drain gap and perhaps even narrow the channel away from the well-edge, e.g., shape it more like an Aharanov-Bohm ring~\cite{WebbPRL85} than a rectangle with a hole, although this would carry more complex fabrication and/or lower device yield. As a final note however, very long pre-assay hydration protocols, e.g., overnight or longer, proved to be counterproductive. At this time-scale, water begins to creep under the CYTOP layer causing delamination at the base of some fraction of the microwells in the array.

As additional data, Supplementary Figure~S10 shows control data for our experiment comparing three devices from Figure~5 with three devices from a separate assay where the $\alpha$-hemolysin concentration of the $\alpha$-hemolysin solution is zero to demonstrate a baseline for change in fluorescence and conductance in a bilayer-sealed device microwell. Supplementary Figure~S11 shows fluorescence intensity and OECT conductance data for four selected devices on two separate coverslips performed as separate assays. Supplementary Figure~S12 shows a single device trace from each of six different coverslips/assays, confirming that the essential aspects of our experiment are reproducible/repeatable, albeit with some variation in the signal magnitude, particularly in the OECT conductance from assay-to-assay. Although we typically obtain $>90\%$ of devices showing measurable ion-sensitivity on our coverslips, under real assay conditions such as those in Fig.~5 we typically get useable data from far fewer OECT microwells. In our experience so far, this is at best $15-20$~devices and sometimes fewer than $5$. Much of the yield loss here comes from microwells that fail to seal during the aqueous-organic-aqueous exchange, which occurs more frequently for the device microwells than non-device microwells, presumably because the hydration dynamics of the PEDOT:PSS exerts a hydrostatic pressure against the lipid bilayer sealing the microwell. The remaining yield loss arises from microwells where the lipid bilayer ruptures during the assay, which is likely an outcome of using $\alpha$-hemolysin as the membrane protein model for our experiment. The choice of $\alpha$-hemolysin is a double-edged sword in this respect -- on the one hand it presents a robust, reliable and well-characterized membrane protein that is easy to insert into a lipid bilayer, but on the other it tends to destabilise the bilayer making it more susceptible to rupture. Our approach here is readily extendable to other membrane proteins, and could be taken to the few or even single molecule per microwell limit, as shown by Watanabe {\it et al.}~\cite{WatanabeNatComm14}. However, given that Poisson statistics in this limit results in the largest number of microwells having no membrane proteins inserted, to achieve this limit, one would need to continue to increase the number of OECTs per coverslip from the $52$ we have presently to hundreds, thousands or even tens of thousands. This is entirely possible using the fabrication approach we have deployed in our work. The challenge will be in dealing with the electrical interface between the coverslip itself and the external measurement circuitry, which will be a subject of future development of this device platform.

\section*{Discussion}
We have reported a method for producing an array of fifty two ion-sensitive PEDOT:PSS organic electrochemical transistors on glass coverslip, each with its own fluidically-independent microwell sealed with lipid bilayer, into which membrane proteins can be inserted for simultaneous electrical and fluorescence microscopy studies. To demonstrate the capability of our device, we fill each microwell with an `inner' phosphate assay buffer solution containing $20~\mu$M Alexa-488 dye and $50$~mM KCl, seal with lipid bilayer using an aqueous-organic-aqueous liquid exchange technique~\cite{WatanabeNatComm14}, and fill the flow-cell with a common `outer' phosphate assay buffer that is dye-free and contains $100$~mM KCl. We then insert $\alpha$-hemolysin, which embeds into the lipid bilayer and forms a heptameric pore~\cite{WalkerJBiolChem92, GouauxPNAS94} with diameter $\sim 2.6$~nm~\cite{SongSci96}, sufficient for K$^{+}$ ions to diffuse into the microwell and Alexa-488 dye molecules to diffuse out of the microwell. These produce a drop in the conductance of the transistor and the fluorescence intensity of the microwell, respectively, albeit at very different time-scales that are directly related to the size difference between the K$^{+}$ ion and the Alexa-488 molecule. The simultaneous detection allows us to ensure that the electrical signal arises from the membrane protein itself and not, e.g., rupture of the bilayer. Our devices are readily produced using standard photolithographic methods, and we show a new process whereby the hydrophobic fluoropolymer layer required to enable the bilayer seals is first rendered hydrophilic by O$_{2}$ plasma treatment to enable photolithography using standard resist formulations and then restored to full hydrophobicity using SF$_{6}$ plasma treatment. This removes the need to use highly viscous photoresist for patterning of the fluoropolymer as in prior work~\cite{WatanabeNatComm14}. The ability to use standard microfabrication techniques to pattern microwell arrays with PEDOT:PSS OECTs incorporated into the bottom of them is both scalable and versatile, with larger arrays and the study of other membrane proteins, incorporated either directly or via proteoliposome fusion~\cite{WatanabeNatComm14}, expected to be possible and a focus of future work. One key reason to increase the number of OECTs per coverslip is to facilitate counting of the number of membrane proteins per microwell from a statistical analysis of the conductance data. An increase in OECTs per coverslip by an order of magnitude or two should be sufficient to enable this. Our structures offer a pathway to integrated high-density membrane-protein-augmented transistor arrays for applications ranging from chemometric sensing to neuromorphic computing. For example, our development here would enable complex circuit structures that exploit the slow kinetics and ion-to-electron transduction capacity of PEDOT:PSS OECTs~\cite{vandeBurgtNatElec18} for circuits underneath the fluoropolymer, with the ability to interface some or all of the OECTs with a `front end' that presents a `neuron-like' bilayer interface hosting ligand-gated, light-gated or voltage-gated ion channels~\cite{RivnaySciAdv17}. Our work also highlights the promising scaling and integration prospects for advanced microfabrication of PEDOT:PSS OECT arrays. 

\bibliography{Kojima_ArXiv26_V3}
\bibliographystyle{sciencemag}

\section*{Acknowledgments}
The authors thank Naoki Soga, Shingo Honda, Paul Meredith, Damia Mawad, Lorenzo Travaglini, Siddharth Doshi, Kate Michie, Natalie Plank and Katherina Gaus for helpful discussions during the project. APM thanks Sam Shelton and Doohan Murphy for assistance with hardware development, Roman Lyttleton for assistance with fabrication development, and Jakob Seidl for assistance with software development in the early stages of the project.
\paragraph*{Funding:} This work was funded by the Australian Research Council (ARC) under DP170104024, DP210102085 and DP230102655. APM was a Japan Society for the Promotion of Science (JSPS) Long-term Invitational Fellow during the early development work for this manuscript. The work was performed in part using the NSW node of the Australian National Fabrication Facility (ANFF).
\paragraph*{Author contributions:} APM, HN and LKL conceived the research. SK, MSM and APM designed and/or fabricated the devices. JGG and APM designed and built the electrical measurement apparatus. SR designed and built the fluorescence microscope and microfluidics systems. SK performed the device characterization and ion-sensitivity studies. SK and SR performed the $\alpha$-hemolysin assays and simultaneous electrical/optical measurements. MSM and JGG performed the diffusion rate modelling. SK, SR, APM and LKL contributed to the data analysis. APM wrote the manuscript, and all authors participated in manuscript input and editing.
\paragraph*{Competing interests:} There are no competing interests to declare.
\paragraph*{Data and materials availability:} All data needed to evaluate the conclusions in the paper are present in the paper and/or the Supplementary Materials.

%%%%%%%%%%%%%%%% SUPPLEMENT LIST %%%%%%%%%%%%%%%

\subsection*{Supplementary materials}
Materials and Methods\\
Supplementary Text\\
Figures S1 to S20\\

%%%%%%%%%%%%%%%% END OF MAIN TEXT %%%%%%%%%%%%%%%
\newpage

%%%%%%%%%%%%%%%% START OF SUPPLEMENT %%%%%%%%%%%%%%%

\renewcommand{\thefigure}{S\arabic{figure}}
\renewcommand{\thetable}{S\arabic{table}}
\renewcommand{\theequation}{S\arabic{equation}}
\renewcommand{\thepage}{S\arabic{page}}
\setcounter{figure}{0}
\setcounter{table}{0}
\setcounter{equation}{0}
\setcounter{page}{1}

%%%%%%%%%%%%%%%% SUPPLEMENT TITLE PAGE %%%%%%%%%%%%%%%

\begin{center}
\section*{Supplementary Materials for\\ \scititle}
S.~Kojima, S.~Rawat, M.~Sanchez~Miranda, J.G.~Gluschke, H.~Noji, L.K.~Lee, A.P.~Micolich$^{\ast}$\\
\small$^\ast$Corresponding author. Email: adam.micolich@nanoelectronics.physics.unsw.edu.au
\end{center}

\subsubsection*{This PDF file includes:}
Materials and Methods\\
Supplementary Text\\
Figures S1 to S20\\

\newpage

%%%%%%%%%%%%%%%% MATERIALS AND METHODS %%%%%%%%%%%%%%%
\subsection*{Materials and Methods}
\subsubsection*{Device Fabrication}
We produce our devices on $40 \times 22$~mm $\#1.5$ glass coverslips (Corning), with nominal thickness $0.17$~mm, which are cleaned using a $1\%$ aqueous solution of Hellmanex-III detergent (Merck) in an ultrasonic cleaner for $10$~minutes, prior to a brief ethanol rinse and dry with an N$_{2}$ jet. In the first fabrication stage, we define the metal layer as follows. We expose the coverslips to an O$_{2}$ plasma at $340$~mTorr and $50$~W microwave power (Denton PE-250 Barrel Asher) for $3$~min to make the surface hydrophilic, followed by a $3$~min dehydration bake at $180^{\circ}$C. To minimise edge-beads during spin-coating of the photoresist, we attach the coverslip to a clean $2"$ silicon wafer~\cite{LuurtsemaMScThesis97,YanCoatings21} using a $\sim 3 \times 8$~mm strip of thin PDMS film. The PDMS film is produced by mixing and de-airing Sylgard 184 (Dow-Corning) before adding $\sim3$~g of the pre-cured PDMS to an $85$~mm diameter plastic petri dish and spinning at $250$~rpm for $15$~s followed by $500$~rpm for $15$~s to spread the film, and then leaving it on a flat surface to cure overnight. We cut the segment of film using a scalpel and use a sharp pair of tweezers to lift it out and place it on the Si wafer before carefully laminating the coverslip on top. We use a dual-layer resist for the metal layer to ensure an undercut edge-profile for clean metal edges after lift-off. We first spin-coat AR-BR-5460/7.8 (AllResist) at $4000$~rpm for $30$~s using dynamic dispense of $350~\mu$L of resist from a pipette (Finnpipette F2). We detach the coverslip from the PDMS (carefully lift from one side using wafer tweezers) to perform a hotplate bake at $150^{\circ}$C for $5$~min before reattaching to the PDMS for the second layer. We then spin-coat ECI-3012 (Microchem) at $4000$~rpm for $30$~s using static dispense of a resist puddle that covers the entire coverslip surface. We once again detach the coverslip from the PDMS and hotplate bake at $90^{\circ}$C for $60$~s. We align and UV expose using a Karl S\"{u}ss MJB-3 aligner to a dose of $65$~mJ/cm$^{2}$ followed by a hotplate bake at $110^{\circ}$C for $90$~s. Development is performed in the TMAH-based AZ2026MIF developer (Microchem) for $90$~s under slight agitation, done in batches using a slide rack (Diversified Biotech WSDR-2000), followed by two rinses in deionized water, before the coverslips are dried with an N$_{2}$ jet. A descum is performed at $340$~mTorr O$_{2}$ at $50$~W (Denton PE-250) for $2$~min prior to vacuum thermal evaporation of $15$~nm Al and $70$~nm Au. This stage is completed with a $30$~min lift-off in acetone (ultrasonic can be used if needed, but use sparingly), followed with an ethanol rinse and dry under N$_{2}$ jet. Samples are checked for quality at this stage and only those with near-perfect metal pass on to subsequent stages. The metal layer contains the transistor contacts, interconnects, pads for external circuit connections, fiducials for alignment of subsequent stages, alignment markers for the glass-block used in assays (see later) and a large `UP' marker (see Fig.~1c), which is extremely useful for ensuring subsequent fabrication steps are all performed on the correct side of the coverslip.

In the second fabrication stage, we define the PEDOT:PSS channels for the OECTs as follows. We start by exposing the coverslips to O$_{2}$ plasma at $340$~mTorr and $50$~W (Denton PE-250) for $10$~min to render the surface hydrophilic and remove any organic contaminants remaining after the preceding stage. We then attach the coverslip to a $2"$ Si wafer using PDMS, as previously described, and spin-coat the PEDOT:PSS mixture (recipe in Materials and Chemicals below) at $950$~rpm for $30$~s (static dispense from a puddle covering whole surface) followed by $4500$~rpm for $1$~s to prevent PEDOT:PSS residue from shorting the contact pads near the four corners of the completed device. This typically produces a PEDOT:PSS layer with thickness $\sim100~$nm. We detach the coverslip from the PDMS and hotplate bake at $140^{\circ}$C for $1$~hr before leaving the coverslip in deionized water overnight to remove low molecular weight components from the PEDOT:PSS film. The sample is rinsed with ethanol and dried with an N$_{2}$ prior to spin-coating AZ-nLOF-2020 resist (Microchem) at $4000$~rpm for $30$~s; attachment to a $2"$ Si wafer for this step is optional. The coverslip is baked at $110^{\circ}$C for $2$~min, before aligning and UV exposing (Karl S\"{u}ss MJB-3) to a dose of $45$~mJ/cm$^{2}$, followed by another hotplate bake at $110^{\circ}$C for $100$~s. We batch develop, as described earlier, in AZ2026MIF developer (Microchem) for $100$~s with slight agitation followed by two rinses in deionized water and drying with an N$_{2}$ jet. The resist is used as an etch-mask for a $\sim30$~min O$_{2}$ plasma etch at $130$~mTorr and $100$~W (Denton PE-250). We confirm completion of the etch by using a multimeter (Fluke $115$) to check there is no measurable resistivity between two points on the glass several mm apart (ideally locations upper left and lower right of the central region in Fig.~1c). If there is any measurable resistivity, we continue the etch for another $5-10$~min and check again. After etching, we lift-off the resist by immersion in acetone at $40^{\circ}$C for $10$~min (this can be done at room temperature but it takes longer), avoiding ultrasonic unless absolutely necessary. The coverslip is rinsed in ethanol and dried with an N$_{2}$ jet.

In the third fabrication stage, we define the microwells in the fluoropolymer layer as follows. We spin-coat CYTOP CTL-809M (Asahi Glass Company) by spin-coating at $800$~rpm for $10$~s to spread the polymer and $4000$~rpm for $30$~s to cast the film. We use a static dispense of $105~\mu$L from a pipette (Gilson M250E) at the centre of the coverslip to minimise spin wastage because this polymer is very expensive. This is followed by two hotplate bakes, at $90^{\circ}$C for $10$~min and $180^{\circ}$C for $30$~min, to cure the film. We expose the film to O$_{2}$ plasma at $340$~mTorr and $50$~W for $2$~min to make the CYTOP surface hydrophilic (contact angle $95^{\circ}$ or less) prior to spin-coating ECI-3012 photoresist (Microchem) at $4000$~rpm for $30$~s with static dispense from a puddle that covers the entire coverslip surface. Attachment to a $2"$ Si wafer is optional; we usually do not do so for either spins in this stage. We hotplate bake the coverslip at $90^{\circ}$C for $60$~s, align and UV expose to $65$~mJ/cm$^{2}$ dose (Karl S\"{u}ss MJB-3), followed by another hotplate bake at $110^{\circ}$C for $90$~s. We batch develop, as described earlier, in AZ2026MIF developer (Microchem) for $60$~s with slight agitation followed by two rinses in deionized water and drying with an N$_{2}$ jet. The resist masks an O$_{2}$ reactive-ion etch performed in a custom parallel-plate RIE system at $75$~mTorr with $20$~sccm O$_{2}$ flow and $50$~W microwave power for $5$~min. This step must be performed in parallel-plate configuration to obtain vertical walls on the microwells and the etch time is set based on etch-rate studies with the aim of exposing bare glass at the bottom of the microwells. This also ensures that the $54$ contact pads at the two ends of the coverslip are properly exposed for electrical contact to the transistors to be obtained. The photoresist is removed by immersion in acetone for $10$~min followed by an ethanol rinse and drying with an N$_{2}$ jet. The final step is an SF$_{6}$ plasma treatment to restore the hydrophobicity of the CYTOP surface, this can be performed during the fabrication process, but we often delay this step until a day or two before the device is actually used in an assay to avoid degradation of the CYTOP surface hydrophobicity during storage from impacting our yield of lipid-sealed microwells. This step is performed in a custom parallel-plate RIE system at $20$~mTorr with $20$~sccm SF$_{6}$ flow and $120$~W microwave power for $5$~s aiming to restore the CYTOP to its `as-spun' hydrophobicity (contact angle $\sim110^{\circ}$).

For some studies in this paper, we use slides where the metal and PEDOT:PSS layers are omitted, which we refer to as `Tokyo' slides since they are essentially the same devices used by the Noji lab in their earlier studies (notably Watanabe {\it et al.}~\cite{WatanabeNatComm14}). These coverslips are made by first cleaning using a $1\%$ aqueous solution of Hellmanex-III detergent (Merck) in an ultrasonic cleaner for $10$~minutes, prior to a brief ethanol rinse and dry with an N$_{2}$ jet, followed by the third stage described above. The only modification is the photomask itself, which has the complete array of microwells -- none of the microwells are removed as there are no metal/PEDOT:PSS layers underneath that need to be insulated -- and no fiducials for alignment to the metal stage, we simply align the mask using the coverslip edges.

We find an essential component of the fabrication to be that the pattern is located at the centre of the mask (we use $4"$ square chrome on soda-lime glass masks), and thus we use four separate photomasks for our device fabrication. The GDSII files for each of these masks are available for download at https://github.com/AdMico/LipidWells. An overlay of the three masks used for our devices is also provided in Supplementary Figure~S13.

\subsubsection*{Materials and Chemicals}
{\bf Fabrication:} All resists and polymers listed above are used as supplied unless specified otherwise. We use a PEDOT:PSS mixture for fabrication of our OECTs that is made as follows. The composition of the mixture is nominally $93.75\%$ PH1000-grade PEDOT:PSS (Ossila), $5\%$ ethylene glycol (Merck), $0.25\%$ dodecylbenzene sulfonic acid (Merck) and $1\%$ (3-glycidyloxypropyl)-trimethoxysilane (Merck). Since there is little in the literature about how to prepare this mixture for use, we give our full recipe here. First, we prepare a $20:1$ stock mixture of EG:DBSA because the small fraction and extremely high viscosity of DBSA makes it very difficult to accurately add it as an individual component. We prepare this in a $50$~mL Falcon tube to which we add approximately $2$~mL of DBSA and $40$~mL of EG. We add the DBSA first, noting as accurately as possible how much we add (working by mass is most accurate), and tailor our quantity of EG accordingly. We stir the mixture vigorously using magnetic stirring and/or vortexing to assure a thorough mixing of the components prior to use. Next, we decant GOPS from the stock bottle into $10$~mL Headspace septum vials (Supelco) in an N$_{2}$ glovebox. We sonicate the Clevios PH1000 PEDOT:PSS (Ossila) stock bottle for $15-30$ minutes before use to obtain a uniform suspension. We typically prepare the PEDOT:PSS mixture in a quantity sufficient for a batch of six devices as follows. We add $3$~mL PH1000 and $168~\mu$L of the EG:DBSA mixture to a $5$~mL eppendorf tube and briefly vortex. We then pull $32~\mu$L of GOPS using a glass Hamilton syringe (Merck), add to the mixture and vortex again. Finally, we syringe filter the mixture into a second Eppendorf tube using a $2$~mL Luer-lock syringe with a $0.45~\mu$m PES filter (Ossila), this sometimes requires more than one filter as they block with particulates somewhat rapidly. There is a $\sim30\%$ loss to filtration giving $2.2$~mL of final mixture, sufficient for six dispenses of $350~\mu$L for spin-coating. Any excess is discarded, it cannot be used later due to the cross-linker, hence the preparation of small batches for each fabrication run.

{\bf Assay Solutions:} All ion-sensitivity studies are performed using KCl solution. We start with a $1$~M KCl stock and prepare solutions at $100$~mM, $10$~mM and $1$~mM by sequential dilution (1:9 in deionized water). Any aqueous solution used in assays is based on a Phosphate Assay Buffer (PAB) made as follows. We prepare $1$~M KH$_{2}$PO$_{4}$, K$_{2}$HPO$_{4}$, KCl and MgCl$_{2}$ stock solutions by adding salt (Merck) to deionized water. We prepare $0.1$~M phosphate buffer as 1:5:54 KH$_{2}$PO$_{4}$:K$_{2}$HPO$_{4}$:H$_{2}$O using the $1$~M stock solutions and deionized water, and then titrate to pH $7.5$ using $1$~M NaOH or $1$~M HCl as appropriate.

For the aqueous components of the assays, we make two different PAB versions, one with $100$~mM KCl and one with $50$~mM KCl. The former is 50:1:5:444 KCl:MgCl$_{2}$:phosphate:H$_{2}$O and the latter is 25:1:5:469 KCl:MgCl$_{2}$:phosphate:H$_{2}$O using the stock solutions above in deionized water. $\alpha$-hemolysin was obtained as lyophilised powder from Creative Enzymes. We added the $500~\mu$g of powder to $1$~mL of $100$~mM KCl PAB, which was divided into $5~\mu$L aliquots at $500~\mu$g/mL ([$\alpha$-hemolysin]~$\sim~15~\mu$M), and stored at $-80^{\circ}$C. Alexa 488 dye (Thermofisher) was diluted into $5~\mu$L aliquots at $2$~mM dye concentration in PAB. The inner solution for our assays is made by adding two $5~\mu$L aliquots of dye to $990~\mu$L of $50$~mM KCl PAB per mL of inner solution for a final dye concentration of $10~\mu$M. The outer solution for our assays is dye-free $100$~mM KCl PAB. For the $\alpha$-hemolysin incubation stage, we use a modified outer solution consisting of two $5~\mu$L aliquots of $\alpha$-hemolysin stock diluted in $990~\mu$L of $100$~mM KCl PAB per mL of outer solution ([$\alpha$-hemolysin]~$\sim~150$~nM or $5~\mu$g/L).

For the organic component of the assay, we start from glass vials of 1,2-dioleoyl-{\it sn}-glycero-3-phosphoethanolamine (DOPE) and 1,2-dioleoyl-{\it sn}-glycero-3-phosphoglycerol (DOPG), each as a solution at $100$~mg in $4$mL of chloroform (Avanti Polar Lipids). We prepare our stock of $4$~mg/mL 1:1 DOPE:DOPG by transferring $160~\mu$L of each lipid above to an $4$~mL brown glass vial (Agilent), and then evaporating off the chloroform with N$_{2}$ gas. The lipid is stored dry under N$_{2}$ atmosphere in the sealed vials at $-80^{\circ}$C until needed for use. The lipid in a given vial is reconstituted as needed by adding $2$~mL of chloroform (Aldrich) and $2.5~\mu$L of deionized water. This small amount of added water, close to the miscibility limit of water in chloroform, assists with the stability of the bilayer by reducing the strong polar-nonpolar stress on the lipid molecules in the border region between the outer leaflet, inner leaflet and upper rim of the CYTOP well (also known as the Plateau-Gibbs border).

\subsubsection*{Electrical Measurements}
A completed coverslip is mounted in a custom-built device mounting system, which accurately locates a set of $54$ Harwin spring pins (Mouser S13-503/P13-4023) to make reliable contact to the $54$ large contact pads ($2\times1$~mm) on the coverslip (see Supplementary Figure~S14). The $54$ Harwin pins are soldered to a custom PCB which passes them into 30-line flexible ribbon cables (Samtec/Mouser) for connection to our custom-built multiplexer box, which manages the connections of the various device lines to the two Femto DLPCA-200 current pre-amplifiers and one National Instruments USB-6216-BNC digital-analog converter system used in our measurement system (see Supplementary Figure~S15). Our multiplexer system consists of a set of four Analog Devices MAX306 16-channel CMOS analog multiplexer chips and associated control/power circuitry, and is driven by the GPIO outputs of a Raspberry Pi 4b using custom python software.

Electrical measurements are performed by applying a d.c. source-drain voltage $V_{sd}$ on the common source line and measuring the d.c. current $I_{d}$ on a given device's drain line. Our configuration is designed so that we can measure two of the $52$ OECTs simultaneously, one from the left branch of devices (OECTs $1-27$) and one from the right branch of devices (OECTs $28-52$). The connections to the respective devices are managed by the custom-built multiplexer system, which feeds the two returning $I_{d}$ signals at a given point in time to the pair of current pre-amplifiers set to a gain of $10^{4}$~V/A. The USB-6216-BNC is used to read out the two current pre-amplifier outputs via Analog Inputs 0 and 1, supply $V_{sd}$ via Analog Output 0, and also supply any required gate voltage $V_{g}$ or simply hold the Ag/AgCl reference electrode at $0$~V using Analog Output 1. The measurement is controlled by a PC running custom Python code, which communicates with the USB-6216-BNC via the \textsc{nidaqmx} python library and a Raspberry Pi 4b, via the \textsc{gpiozero} python library. Measurements are sometimes obtained `single-shot' by polling the analog inputs of the USB-6216-BNC for a single value, but more often obtained as `burst' measurements by polling the analog inputs at $400$~kHZ for a short period to obtain a sample of values, from which we obtain the average (signal) and standard deviation (noise/uncertainty). The sample count per burst is typically $1000$, which is the smallest value where obtaining more samples yields no tangible improvement in the uncertainty of the measurement. Our software then runs the measurements as a discrete number of `grabs', each of which is a read of all $52$ OECTs, before a timed pause preceding the next grab. A typical grab takes $\sim1$~s with about $10$~s pause between each grab.

The part and PCB designs for our sample mounts and multiplexer circuit are available at https://github.com/AdMico/LipidWells and our custom python software is available at\\ https://github.com/AdMico/PyNE-wells.

\subsubsection*{Microfluidics and Fluorescence Microscopy}
We take completed device coverslips and make a closed flow-cell on them as follows (see Supplementary Figure~S16). A $20\times15\times6$~mm BK7 glass block (Eksma Optics) with two $\sim1-2$~mm diameter conical holes drilled for the inlet and outlet lines is coating on the bottom-side with CYTOP. We do this by pipetting $60~\mu$L of CYTOP (CTL-107MK) onto the glass block, spreading it over the entire block-surface with a lint-free tip, and then spinning at $500$~rpm for $30$~s to remove the excess. We then bake the glass block for $10$~min at $90^{\circ}$C and $30$~min at $180^{\circ}$C to cure the CYTOP film. This coating faces downwards in the assay configuration presenting a hydrophobic top surface to match the hydrophobic surface of the coverslip. The flow-cell is held together using a Protease Chain Reaction (PCR) frame-seal (BioRad SLF0201) with a $9\times9$~mm inner opening area and a height $0.3$~mm (nominal total volume $25~\mu$L). The frame-seal and glass block are carefully aligned to a set of metal markers on the device coverslip to ensure reliable placement of the flow-cell in each experiment.

The design of the flow-cell makes the addition of an Ag/AgCl gate/reference electrode challenging. We previously had a third hole drilled into the glass block through which we passed an Ag/AgCl microelectrode (InVivoMetric E215P), however this presented unresolvable issues with air-bubble formation inside the flow-cell. Instead we use an $300~\mu$m diameter Ag wire (ESPI Metals), which we immerse in $6\%$ w/w sodium hypochlorite solution for 1~hour, rinse with water, and then thread into the flow-cell as follows. First, we insert a pair of pipette tips into the glass block and adhere them into place using UV-curable epoxy (Norland $\#68$). We then thread the microfluidic tubing and Ag wire into a short section of slightly thicker tubing, which we feed into the pipette tip, and then epoxy both tubing joints to seal the system (see Supplementary Figure~S17). The Ag/AgCl wire length was adjusted such that $2-5$~mm of wire protrudes out of the pipette tip into the flow-cell. The outlet is made in a similar manner without the electrode wire or thicker segment of tubing (i.e., tubing goes direct into pipette tip).

An Elveflow microfluidics system consisting of a pressure controller (OB1 Mk~4) and twelve channel fluidic multiplexer was used to manage fluid flow to our flow-cell from four separate reservoirs: inner solution, lipids in chloroform, outer solution and outer solution with $\alpha$-hemolysin (see schematic in Supplementary Figure~S18). A digital flow sensor is put in the fluidic line from the multiplexer to the flow-cell to provide a stable flow rate. We use PTFE tubing ($300~\mu$m inner diameter) because it is inert to chlorinated solvents (e.g., chloroform). The typical flow-rate used was $50~\mu$L/min for aqueous solutions and $30~\mu$L/min for the chloroform solution. Our microfluidic system has a dead volume of $\sim85~\mu$L. All fluid reservoirs except for the outer solution with $\alpha$-hemolysin are kept immersed in crushed ice during the assay. This is particularly important for the lipid/chloroform solution; lowering the temperature suppresses evaporation leading to a smoother lamellar flow and stabilizes lipid–solvent interactions reducing interfacial instabilities, giving a higher yield of sealed wells and better reproducibility from assay to assay.

A custom-built dual-wavelength fluorescence microscope was used to image fluid-wells in parallel with the electrical measurements (see Supplementary Figure~S19). The first excitation line features a $800$~mW LED at $470$~nm (Thorlabs M470L5) paired with a $495$~nm dichroic filter (Chroma ET495) and used for excitation of the Alexa-488 dye. The second excitation line features a $940$~mW LED at $660$~nm (Thorlabs M470L5) paired with a $645$~nm dichroic filter (Chroma ET645) and used for excitation of Alexa-648 dye, which we use for diagnostics only because it is too large to diffuse through the $\alpha$-hemolysin pore. The two excitation beams pass through a dichroic beam combiner (Nikon DM500) before being collimated using a beam-expander configuration consisting of a $100$~mm focal length collimating lens and a $10\times$ objective lens (Nikon Plan-NEOFLUAR), giving a nearly collimated beam at the output of the objective lens. The emission signal returning from the objective is split off by a quad dichroic mirror (Semrock BrightLine 405/488/561/635~nm) placed between the objective lens and the collimating lens. The beam then passes through an $525$~nm or $690$~nm dichroic filter (Chroma ET525/ET690) mounted on a motorized slider followed by a $200$~mm focal length tube lens. The tube lens image is projected onto the CMOS sensor (ZWO ASI2600MM Pro with $6248\times4176$~pixel CMOS sensor). The device coverslip is mounted in a custom-built microscope sub-stage mounted on a Mad City Labs two-axis servo-controlled stage used to translate the sample along the $x$-$y$ plane using a controller (MCL Micro-Drive). For focus control, the objective is mounted on an L-shaped bracket connected to a vertical translation stage driven by a DC servo motor (Thorlabs MT1/Z825B/KDC101) to enable computer-controlled fine-focus adjustment during the assays. During assays, our main interest is the region containing the OECTs at the centre of the coverslip (see Supplementary Figure~S20). We have written custom python code to auto-focus on and acquire fluorescence images of a $1214\times1214~\mu$m field of view containing all $52$ OECTs that runs in parallel with the software obtaining the electrical measurements such that the intensity and conductance data have common time-stamping for use in the later analysis.

\subsubsection*{Implementation of the $\alpha$-hemolysin Assay}
Once the flow-cell is assembled and the microfluidic lines are attached, the device is mounted on the custom-built microscope sub-stage and held in place with the spring-pin assembly. The Ag/AgCl electrode is held at $0.0$~V and the other device lines are all held at ground potential. We commence the assay by flowing inner solution (phosphate assay buffer with $20~\mu$M Alexa-488 and $50$~mM KCl) to fill the flow-cell. The device coverslip, along with the microscope sub-stage, are mounted in a special aluminium holder sitting in a container of crushed ice to remove any trapped air and ensure all wells are completely filled with inner solution. The device is left on ice for several hours to enable the PEDOT:PSS OECTs to hydrate. Next, we flow a 1:1 w/w solution of DOPE:DOPG at $4$~mg/mL in chloroform to fill the flow-cell to form the inner leaflet of the lipid bilayer sealing each microwell. We immediately flow outer solution (dye-free phosphate assay buffer with $100$~mM KCl) to fill the flow-cell to form the outer leaflet of the lipid bilayer sealing each microwell. Once the aqueous-organic-aqueous exchange stage is complete, the coverslip/sub-stage is returned to the microscope and we allow it to warm to room temperature and stabilise for $\sim10$~min before a short flush with outer solution to remove any dye that escaped from microwells that ruptured during the equilibration process. During this period, we monitor the OECT conductances and fluorescence intensity of wells in the region of interest (ROI), and once we are sure simultaneous recordings are being registered correctly, we commence the flow of the outer solution containing $5~\mu$g/mL $\alpha$-hemolysin at $50~\mu$L/min. Once the flow-cell is filled, we continue the flow for $3-5$~min at $30~\mu$L/min and then stop the flow for the remainder of the experiment. We consider $t = 0$ to be the point where we initiate the flow of hemolysin-containing outer solution and typically run the measurement for $1$~hour afterwards to gather the experimental dataset.

\subsubsection*{Fluorescence Microscopy Image analysis}
All image analysis was performed in \textsc{ImageJ} using a macro we developed to automate the analysis of the sequence of images and obtain the fluorescence intensity versus time for each microwell in the $1214\times1214~\mu$m field of view containing the OECTs. A typical image is shown in Supplementary Figure~S17. The macro operates as follows. First, the acquired 16-bit images were imported into \textsc{ImageJ} as an `image stack', with the sequence sorted automatically using the time-stamp assigned to the image during data acquisition. We then use the \textsc{StackReg} plug-in (Philippe Th\'evenaz, EPFL) to correct for any lateral drift in the image stack. We finish pre-processing of the stack by correcting for any variations in background intensity using the background subtraction algorithm built into \textsc{ImageJ} and applying a Gaussian blur (with $\sigma~=~2$) to reduce noise and enhance the automated microwell detection that follows (i.e., by reducing risk of noise generating false microwells). The following processing steps are done with an `image margin' in place to avoid clipped or optically-distorted wells at the edges of the field-of-view from contaminating the data. First, our macro uses the `Find Maxima' function to identify the bright spots in each image in the stack, which are predominantly microwells filled with dye. We then generate a set of circular `regions of interest' (ROI) with a diameter of $20$~pixels (i.e., expected diameter of our microwells) centred on the maximum for each spot. These ROIs are allocated a number that remains fixed through the stack and is stored in the ROI Manager in \textsc{ImageJ}. The macro then generates a labelled image from the first frame with all of the detected and analysed microwells numbered accordingly, so we can track the intensity versus time data back to the microwell in the image it originated from. Lastly, the macro calculates the average pixel intensity and its standard deviation for each circular ROI for each image in the stack. From this, the macro builds a table of intensity versus time for each ROI, and this is output into a .csv file with the ROI number in the file name as part of the final dataset from the analysis. This automated workflow ensures consistent, reproducible extraction of fluorescence signals from multiple microwells, greatly reducing manual effort and minimizing errors in the intensity versus time analysis.

%%%%%%%%%%%%%%%% SUPPLEMENTARY TEXT %%%%%%%%%%%%%%%

\subsection*{Supplementary Text}
\subsubsection*{Managing the hydrophobicity of CYTOP}
The main text describes an approach we developed to manage the hydrophobicity of the CYTOP surface to improve our capability for doing lithographic patterning of CYTOP films. Specifically, we use a brief exposure to O$_{2}$ plasma to render the CYTOP surface more hydrophilic to enable good adhesion of standard photoresist formulations to the CYTOP under spin-coating. Figure~S1a shows side-by-side photographs of the results of an attempt to spin-coat the photoresist ECI-3012 (Microchem) onto a coverslip freshly coated with CYTOP (i.e., without O$_{2}$ plasma treatment) on the left and a coverslip coated with CYTOP which has been treated with O$_{2}$ plasma (Denton PE-250, $340$~mTorr, $50$~W, $60$~s). The plasma-treated coverslip has good photoresist adhesion across the entire coverslip. The untreated coverslip has a few small areas of photoresist adhesion, generally close to the edges; the vast majority of the photoresist spins off the coverslip leaving clean CYTOP surface behind.

\begin{figure}
\centering
{\includegraphics[width = 0.8\textwidth]{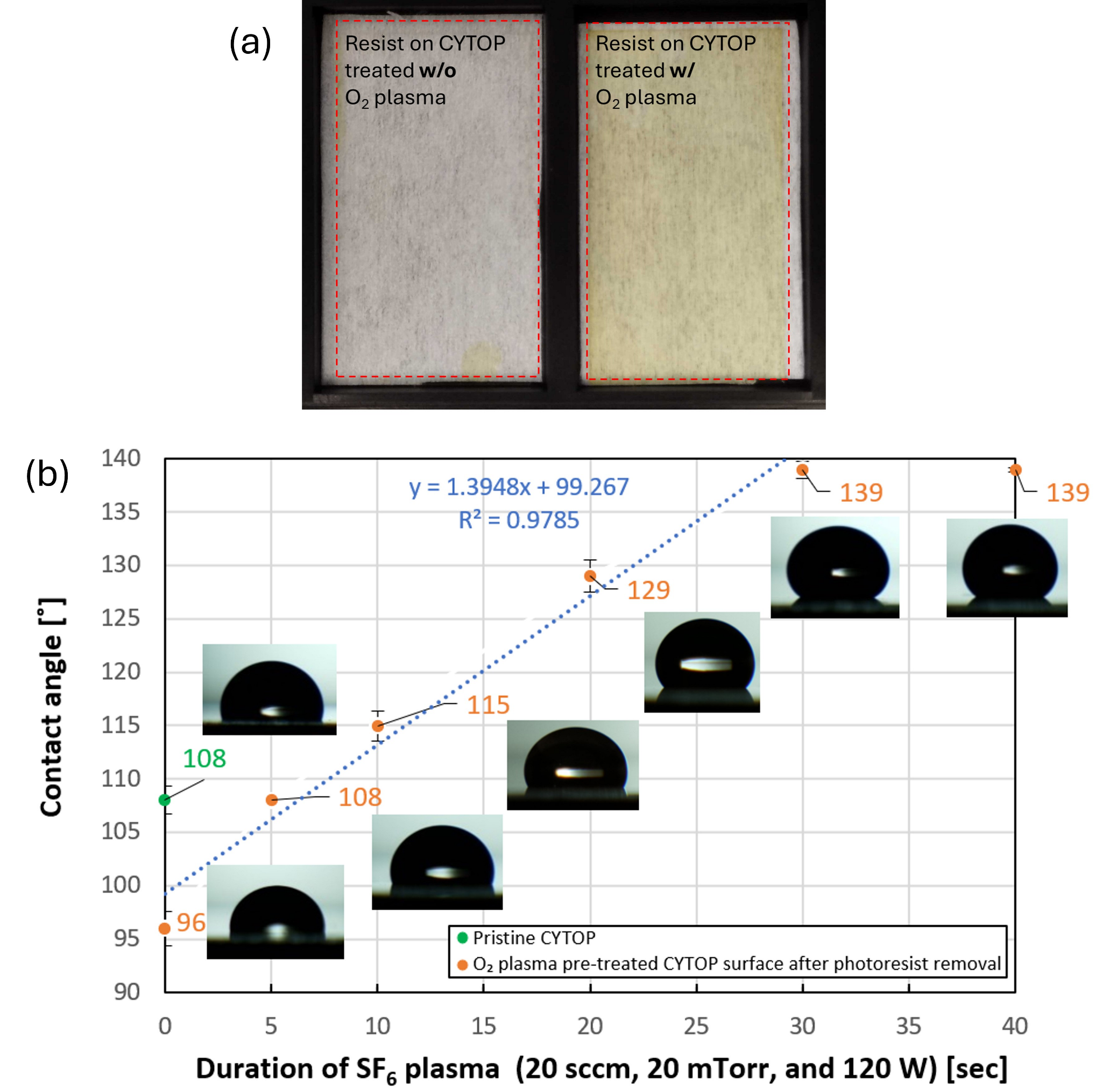}}
\caption{{\bf Control of hydrophobicity of CYTOP via plasma etching.} ({\bf a}) Photographs of two CYTOP-coated coverslips (left) without and (right) with O$_{2}$ plasma treatment prior to spin-coating with ECI-3012 photoresist. Adhesion is extremely poor on untreated CYTOP whilst a $60$~s exposure to O$_{2}$ plasma ($50$~W, $340$~mTorr) renders the CYTOP surface sufficiently hydrophilic that ECI-3012 coats well. This eliminates the need to use very high viscosity photoresists for lithography of the microwells~\cite{WatanabeNatComm14}. ({\bf b}) Plot of the measured H$_{2}$O contact angle vs duration of exposure to SF$_{6}$ plasma ($20$~sccm, $20$~mTorr, $120$~W) for CYTOP-coated glass coverslips rendered hydrophilic by exposure to O$_{2}$ plasma ($60$~s, $50$~W, $340$~mTorr). Data for pristine CYTOP (green) is added for reference. The SF$_{6}$ plasma is able to produce a surface that is significantly more hydrophobic than an as-coated CYTOP surface, and the resulting hydrophobicity can be readily tuned by the etch time.}
\end{figure}

Figure~S1b shows a study of the hydrophobicity of the CYTOP surface as a function of exposure to an O$_{2}$ plasma to make it more hydrophilic for photolithographic patterning and then a SF$_{6}$ plasma to restore the hydrophobicity for other purposes (CYTOP is mostly used to provide high hydrophobicity coatings). The green data point in Fig.~S1b shows the contact angle measured for a pristine CYTOP film on a glass coverslip (CYTOP CTL-809M spin-coated as in the Materials and Methods section). The orange data points are obtained for films that are first exposed to O$_{2}$ plasma (Denton PE-250, $340$~mTorr, $50$~W, $60$~s) and then exposed to SF$_{6}$ plasma for varying times (Custom parallel-plate RIE system, $20$~mTorr, $120$~W, $20$~sccm SF$_{6}$ flow). The O$_{2}$ plasma reduces the water contact angle from $108^{\circ}$ to $96^{\circ}$, and the SF$_{6}$ plasma then progressively increases it back to $108^{\circ}$ after $5$~s and to as high as $139^{\circ}$ after $30$~s. We find that the contact angle saturates after $\sim30$~s, further SF$_{6}$ etching simply reduces the final CYTOP thickness (etch rate is approximately $400$~nm/min).

\begin{figure}
\centering
{\includegraphics[width = 1.0\textwidth]{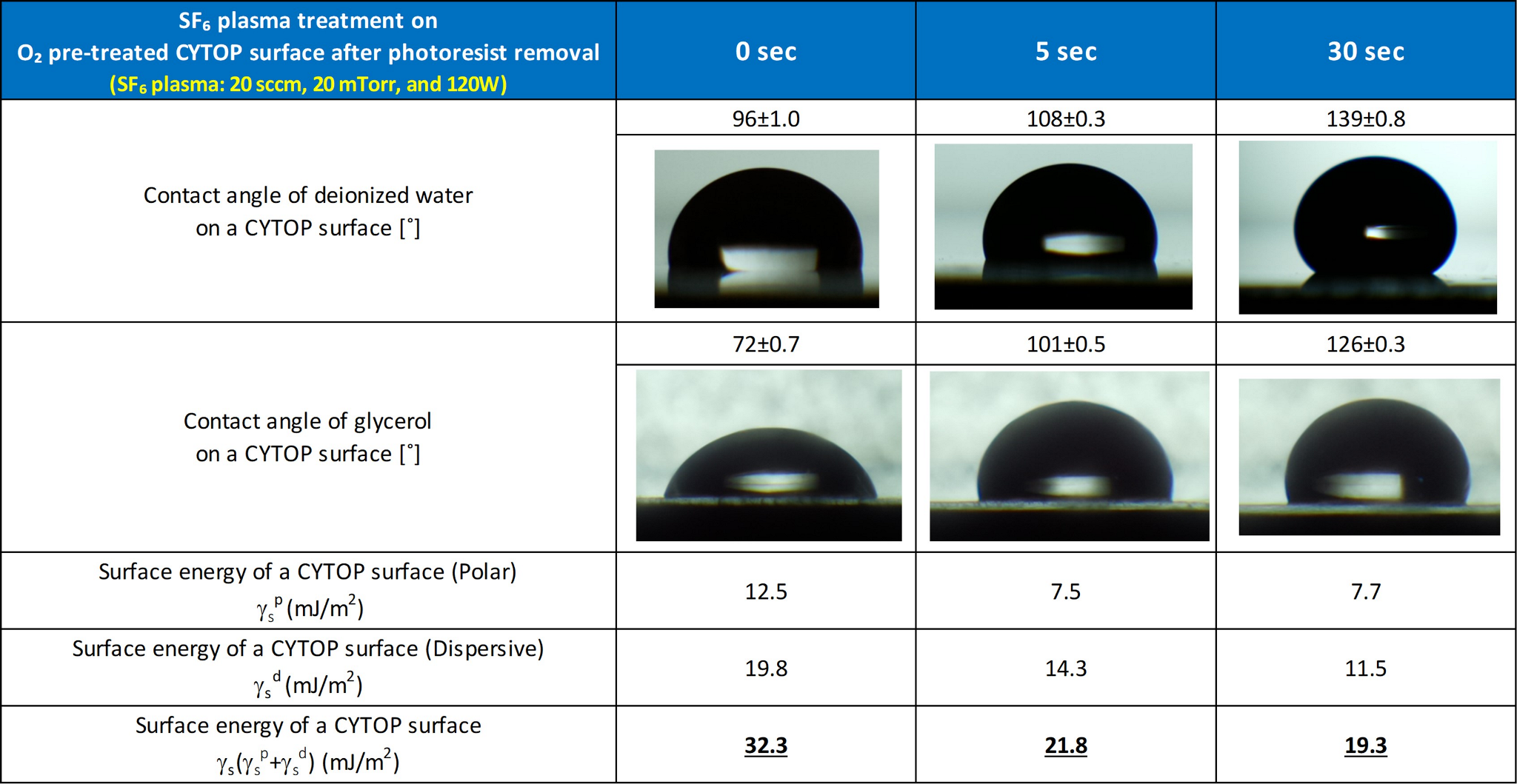}}
\caption{{\bf Surface energy data for SF$_{6}$ plasma processing of O$_{2}$ plasma pre-treated CYTOP.} Results of a study to establish the true surface energy of a CYTOP surface treated first with O$_{2}$ plasma to render the surface hydrophilic and second with SF$_{6}$ plasma to restore the hydrophobicity. This involves performing contact angle measurements with both H$_{2}$O and glycerol~\cite{OwensJAPS69,PelagadeJSEMAT12}.}
\end{figure}

For completeness, Fig.~S2 shows a study of the CYTOP surface energy for the data points at $0$~s, $5$~s and $30$~s in Fig.~S1b, which is obtained by measuring the contact angle of glycerol on the CYTOP surface in addition to the contact angle of water~\cite{OwensJAPS69,PelagadeJSEMAT12}. For reference, the contact angles for water and glycerol on a pristine CYTOP film (i.e., before any plasma treatment) are $108^{\circ}$ and $103^{\circ}$, respectively.

\begin{figure}
\centering
{\includegraphics[width = 1.0\textwidth]{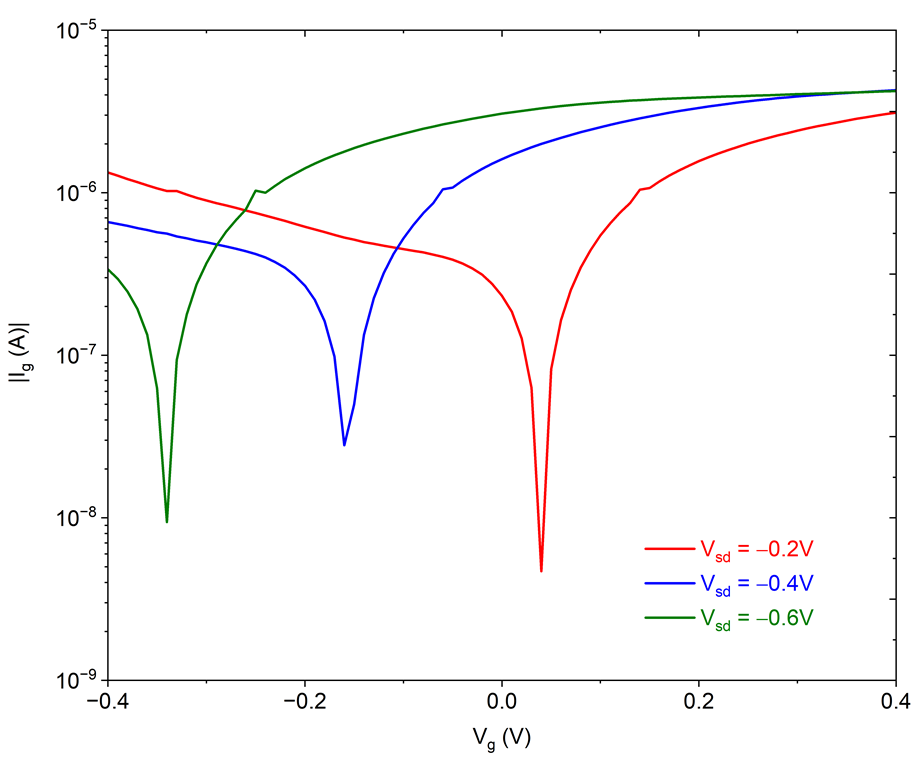}}
\caption{{\bf Measurements of gate electrode current during OECT characterisation.} Plot of the absolute value of the Ag/AgCl gate electrode current $|I_{g}|$ versus the applied gate electrode voltage $V_{g}$ for three different source-drain biases $V_{sd}$ corresponding to those studied in Fig.~2b of the main text.}
\end{figure}

\subsubsection*{Gate current and hysteresis data for our PEDOT:PSS OECTs}
Figure~S3 shows a plot of the current $I_{g}$ flowing to/from the Ag/AgCl gate electrode in response to the applied gate electrode voltage $V_{g}$ for the device characterisation shown in Fig.~2 of the main text. Traces for three different source-drain voltages $V_{sd}$ are presented matching those shown in Fig.~2b. An important detail to note is that our devices are OECTs not FETs. In an FET, the channel current $I_{sd}$ dependence on gate voltage $V_{g}$ can only be trusted as arising from `transistor action' if the current $I_{g}$ leaking from the gate to the channel is comparatively negligible, i.e., if $I_{g}$ is large relative to $I_{sd}$ then $I_{sd}$ is mostly gate leakage not transistor action. OECTs operate very differently to FETs~\cite{RivnayNatRevMater18,ProctorJPolySci16}. In an OECT, the ‘gate charge’ is ionic rather than electronic and the channel current $I_{sd}$ is naturally insulated from any gate current $I_{g}$, i.e., this current is not `leakage’ in the traditional FET sense. Any `leakage’ is intrinsically blocked by the fact that gating is mediated by positive ions, which cannot convert into holes and thereby join the electrical conduction along the conducting polymer chains. In OECTs, the gate electrode current $I_{g}$ associated with gate action simply reflects the migration of solvated positive ions in the electrolyte solution between the Ag/AgCl electrode and the interstitial space between the molecular chains of the conducting polymer, where they act to influence conduction in the organic semiconductor channel.

\begin{figure}
\centering
{\includegraphics[width = 1.0\textwidth]{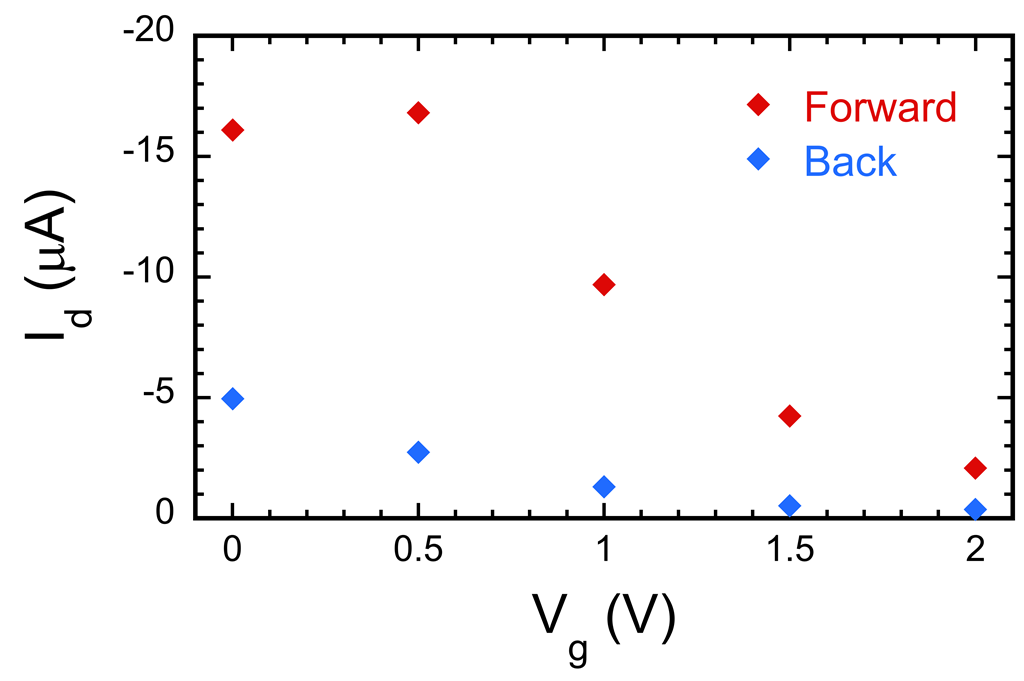}}
\caption{{\bf Preliminary characterisation of gate hysteresis in our OECTs.} Plot of the drain current $I_{d}$ vs gate electrode voltage $V_{g}$ for a forward sweep (red) to positive $V_{g}$ and back sweep (blue) to $V_{g} = 0$. There is a pause of a few minutes between the two traces.}
\end{figure}

The diffusion of ions into and out of the PEDOT:PSS in response to $V_{g}$ is a relatively slow process with an equilibration time that can take minutes, hours and sometimes even days, depending on the volume of the PEDOT:PSS channel, the nature of its interface with the electrolyte, and the internal morphology of the PEDOT:PSS. This leads to well-known hysteresis effects in PEDOT:PSS OECTs~\cite{StavrinidouAdvMat13,KumarAdvElecMat25}, and our devices are no exception. Figure~S4 shows a plot of channel current $I_{d}$ vs gate electrode voltage $V_{g}$ for a forward sweep (red) and a back sweep (blue) with a few minutes pause between the two sweeps. The two traces are substantially different, arising from some combination of hysteresis~\cite{StavrinidouAdvMat13,KumarAdvElecMat25} and channel degradation at positive $V_{g}$~\cite{KeeneMRSComm24}. A large number of cycles under fixed conditions would enable the two contributions to be isolated, but this would depend on the specific conditions, e.g., the $V_{sd}$ and range of $V_{g}$ used. The hysteresis component is likely to be linked to the `long tail' behaviour in some traces in Fig.~5 of the main text and Supplementary Fig.~S9. Channel degradation should not contribute since $V_{g} = 0$ throughout the assay. Establishing the full details of the tail behaviour of our OECT response and its link to hysteresis in PEDOT:PSS requires significant work and will be the subject of future studies.

\begin{figure}
\centering
{\includegraphics[width = 1.0\textwidth]{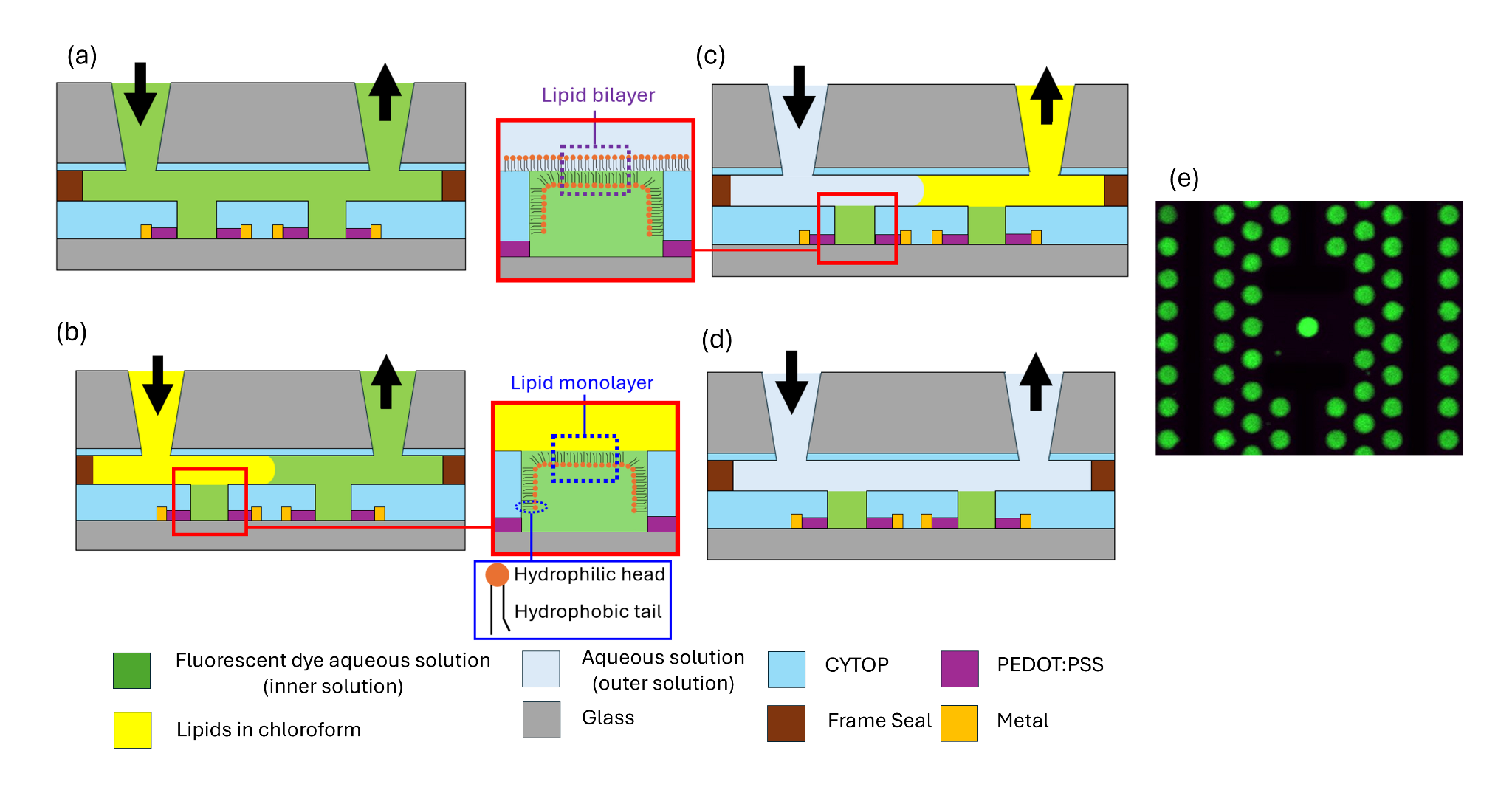}}
\caption{{\bf Schematic diagram of the process for sealing microwells with lipid bilayer.} ({\bf a}) The `inner' solution containing Alexa-488 dye is injected into the flow-cell and fills the microwells. ({\bf b}) A solution of lipids in chloroform is injected into the flow-cell. This displaces the inner solution in the flow-cell but traps inner solution inside the microwells due to the laminar flow. A lipid monolayer forms between the inner solution and the chloroform at the top of the microwell and the inner solution and the CYTOP at the side-walls of the microwell as shown in the associated zoom panel. ({\bf c}) The flow of dye-free `outer' solution flushes the lipid solution out of the flow-cell, driving the formation of an upper leaflet of lipid monolayer across the flow-cell. This results in each microwell being sealed with lipid bilayer such that the inner leaflet is restricted to that microwell whilst the outer leaflet is common across all microwells. The nature of this bilayer means that a membrane protein inserted in a given microwell cannot diffuse beyond that specific microwell. The final state of the system is shown in ({\bf d}) and has the microwell filled with inner solution, and sealed from the outer solution filling the flow-cell by lipid bilayer. ({\bf e}) shows a fluorescence microscopy image of a device coverslip after the process in ({\bf a-d}), the microwells appear bright due to the dye-laden inner solution trapped inside the microwells by the lipid bilayer. The process follows that described by Watanabe {\it et al.}~\cite{WatanabeNatComm14}}.
\end{figure}

\subsubsection*{Mechanism for sealing the CYTOP microwells with lipid bilayer}
We use a similar aqueous/organic/aqueous exchange mechanism to that used by Watanabe {\it et al.}~\cite{WatanabeNatComm14} to seal the microwells with lipid bilayer. We start by assembling the flow-cell and connecting the fluidic lines, as described in the Materials and Methods section. The process that then follows is shown schematically in Fig.~S5. We first fill the flow-cell with the inner solution (phosphate assay buffer with $20~\mu$M Alexa-488 dye and $50$~mM KCl) and place the coverslip onto an Aluminium block sitting in crushed ice. This is a crucial step that `draws' the inner solution into the microwells and purges any trapped air~\cite{WatanabeMMB18}. Successful completion of this step can be determined quite easily. A dry microwell array has an opalescent appearance as it is essentially a photonic crystal. This appearance is retained if the inner solution has not filled the microwells, but vanishes once the microwells are filled because the refractive index of CYTOP is very close to that of water. Typically $3-5$~minutes on the ice is sufficient, however, for our device coverslips, we leave the system in this state for $\sim3-6$ hours to allow the PEDOT:PSS channel material to hydrate. Failing to hydrate the PEDOT:PSS prior to sealing the microwells increases the risk of bilayer rupture and/or device instability due to the water uptake of the PEDOT:PSS competing with the surface tension of the bilayer for the trapped volume of fluid inside the microwell.

The second step is to flow the lipid in chloroform solution ($4$~mg/mL 1:1 DOPE:DOPG) through the flow-cell. The front between the aqueous and organic solutions sweeps past the microwells sealing them with lipid monolayer and leaving the dye-laden inner solution trapped inside the microwells as shown in Fig.~S5b. The third step is to flow the outer solution (dye-free phosphate assay buffer with $100$~mM KCl) through the flow-cell. The front between the organic and aqueous solutions this time seals the microwells with lipid bilayer as it sweeps past, as shown in Fig.~S1c. For this to occur successfully the flows need to be as laminar as possible (minimal turbulence) and have well-controlled and constant flow speed. We use flow speeds of $50~\mu$L/min for the aqueous solutions and $30~\mu$L/min for chloroform solution. The time between the organic flow and the second aqueous flow should ideally be as short as possible.

The final state of the system is shown in Fig.~S5d and features microwells filled with inner solution, sealed with lipid bilayer, and a common reservoir of outer solution in the flow-cell. The geometry of the resulting bilayer, as highlighted in Fig.~S5c, is noteworthy. Each microwell has an independent inner leaflet that runs across the top of the microwell and down the microwell walls due to the hydrophobic CYTOP surface. The outer leaflet is common across large areas of the microwell array and has a corresponding monolayer leaflet across the continuous CYTOP film on the glass block at the top of the flow-cell. To elaborate slightly, a bilayer consists of two leaflets, which we denote as `upper' and `lower'. At the centre of a microwell, those leaflets are very close together and form a true lipid ‘bilayer’ not simply two monolayers in parallel. The proof that this is true lipid bilayer lies in the ability for membrane proteins to embed in and achieve viable catalytic activity at all, i.e., the activity observed in Watanabe {\it et al.}~\cite{WatanabeNatComm14} is impossible unless that seal is a true bilayer. As one approaches the edge of the microwell, there are two possibilities, the upper and lower leaflet stay together, i.e., there is just one single bilayer over everything in the device, or the upper and lower leaflet diverge. If the leaflets diverge, we should be able to see that under a microscope using thin-film interference, and that is indeed the case, as shown in Supplementary Figure~2 of Watanabe {\it et al.}~\cite{WatanabeNatComm14} or Supplementary Figure~2 of Soga {\it et al.}~\cite{SogaSciRep15}. This divergence at the edge of the microwell is formally known as the `Plateau-Gibbs border’~\cite{FujiwaraJChemPhys03,WangAnalyst12}, with the effect very nicely shown in Figure~3 of Wang {\it et al.}~\cite{WangAnalyst12}. What happens after the leaflets diverge at the edge of the microwell can be understood using `asymmetric’ bilayers where one leaflet is has dye-labelled lipid and the other leaflet has unlabelled lipid. This can be achieved by splitting the lipid-in-chloroform step into two steps, one with dye-labelled lipid in the chloroform and the other with unlabelled lipid. If the dye-labelled lipid is used first, the lower leaflet is dye-labelled. If the dye-labelled lipid is used second, the upper leaflet is dye-labelled. A dye-labelled lower leaflet was studied by Watanabe {\it et al.}~\cite{WatanabeSciRep14}. Fluorescence was localised strictly to the microwells showing that the lower leaflet terminates at the inner vertical wall of the microwell. Fluorescence Recovery After Photobleaching (FRAP) studies further confirm that lower leaflet lipid molecules cannot migrate beyond the microwell that they are initially confined to. How far down the microwell wall the lower leaflet extends is difficult to ascertain, but the fact that it at least does can be determined by confocal microscopy. A dye-labelled upper leaflet was also studied by Watanabe {\it et al.}~\cite{WatanabePNAS18}. Fluorescence extended across very large regions encompassing many microwells. Whether that is the entire population of microwells on the coverslip is hard to determine but FRAP studies in this system show that lipid molecules can diffuse over large lateral distances well beyond the microwell they are initially located at. The hydrophobicity of the glass block surface is vital to obtaining sealed microwells. The CYTOP coating on this block is refreshed regularly (often each assay) by either SF$_{6}$ plasma treatment or by recoating with fresh CYTOP. The morphology of the lipid bilayer in our devices means that the microwells are both fluidically-independent and biologically-independent, in the sense that any transmembrane protein inserted into the bilayer cannot diffuse beyond the microwell it occupies. The hydrophilic surfaces at the bottom (i.e., glass and PEDOT:PSS) are in direct contact with the inner solution (i.e., the lipid monolayer only extends where the surface is sufficiently hydrophobic), enabling the PEDOT:PSS OECT to sense the ion concentration inside the microwell, which can be modulated via the activity of any transmembrane proteins inserted into the small patch of bilayer sealing that microwell.

\begin{figure}
\centering
{\includegraphics[width = 0.8\textwidth]{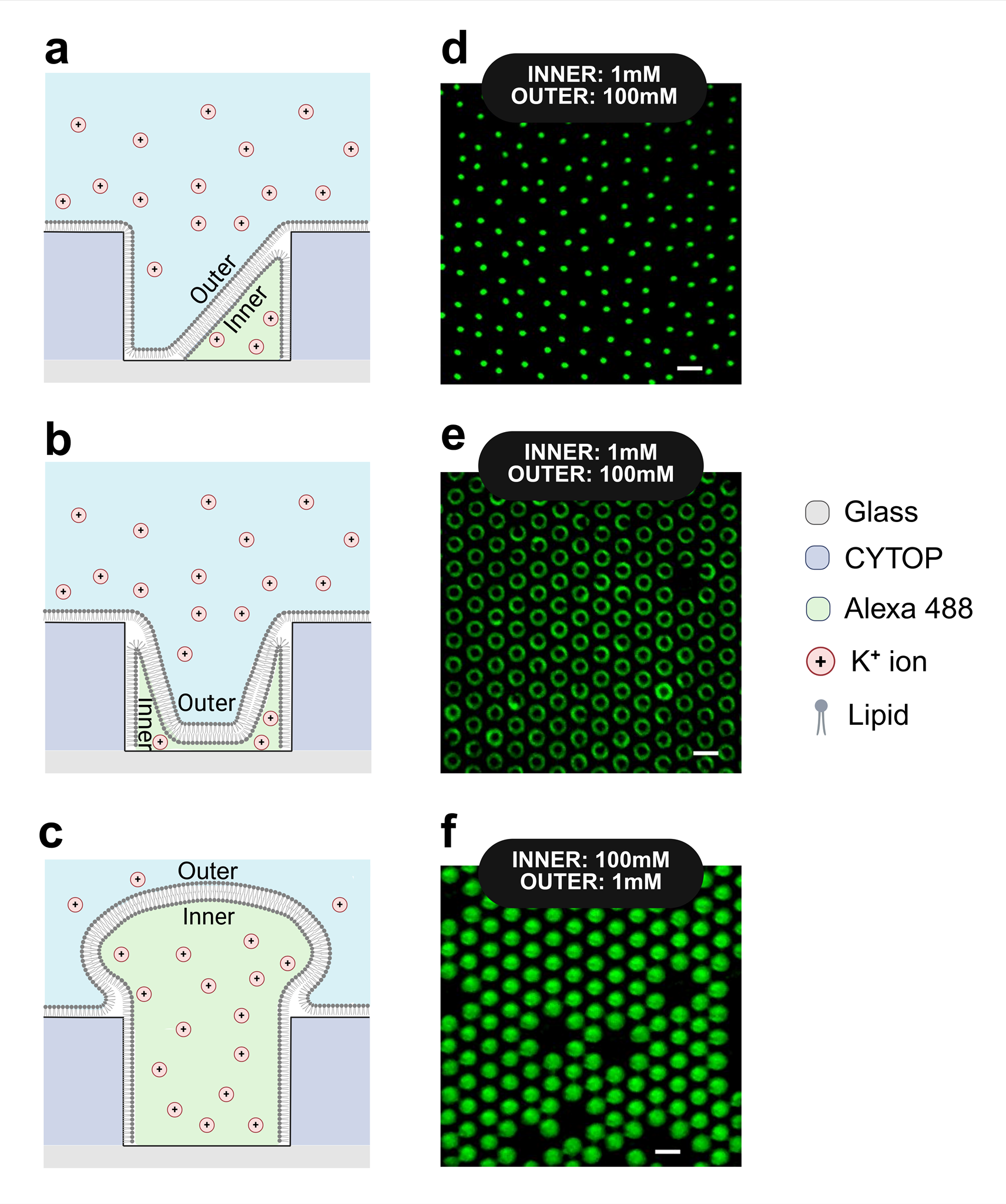}}
\caption{{\bf Bilayer seals under extremes of osmotic pressure.} Schematics of how a large [K$^{+}$] difference across the bilayer affects the resulting shape of the bilayer sealing the microwell for ({\bf a,b}) cases where [K$^{+}$]$_{outer}~\gg~$[K$^{+}$]$_{inner}$ and ({\bf c}) [K$^{+}$]$_{outer}~\gg~$[K$^{+}$]$_{inner}$. ({\bf d-f}) presents corresponding fluorescence images for the three schematics. ({\bf d,e}) were obtained with  [K$^{+}$]$_{outer} = 100$~mM and [K$^{+}$]$_{inner} = 1$~mM, ({\bf f}) was obtained with [K$^{+}$]$_{outer} = 1$~mM and [K$^{+}$]$_{inner} = 100$~mM. In each case, the inner solution has $20~\mu$m Alexa-488 dye and the outer solution is dye-free. The scale bar in ({\bf d-f}) represents $10~\mu$m.}
\end{figure}

\subsubsection*{Bilayer seals under extremes of osmotic pressure}
Differences in the concentration of ions and small molecules between the inner and outer solutions lead to an osmotic pressure gradient across the bilayer sealing the microwell, which in turn causes it to deform outwards or inwards. A larger concentration difference results in a larger deformation of the bilayer. Obtaining a strong electrical signal in our study incentivises increasing the [K$^{+}$] difference across the bilayer, but this can only be done to the extent that the resulting osmotic pressure and bilayer deformation does not adversely impact on the ability to perform the measurement, e.g., by rupturing the bilayer, reducing the volume of inner solution to an impractical extent, or making the fluorescence intensity of the microwell impossible to reliably measure. Figure~S6 shows some examples of this obtained for a [K$^{+}$] difference of two orders of magnitude across the bilayer. For low inner solution [K$^{+}$], the bilayer is pulled inside the microwell, touching either the floor and/or walls, as shown in Fig.~S6a/b. This results in small trapped volumes of inner solution that either only interact with a small area of the available PEDOT:PSS OECT interface (see Fig.~S6d) or are difficult to obtain reliable fluorescence signal from (see Fig.~S6e). For high inner solution [K$^{+}$], the bilayer bows outward, which can move/change shape in the outer solution over time. This also presents difficulty in obtaining reliable fluorescence signal (see Fig.~S6f), and increases the risk of the bilayer rupturing during the measurement.

\begin{figure}
\centering
{\includegraphics[width = 1.0\textwidth]{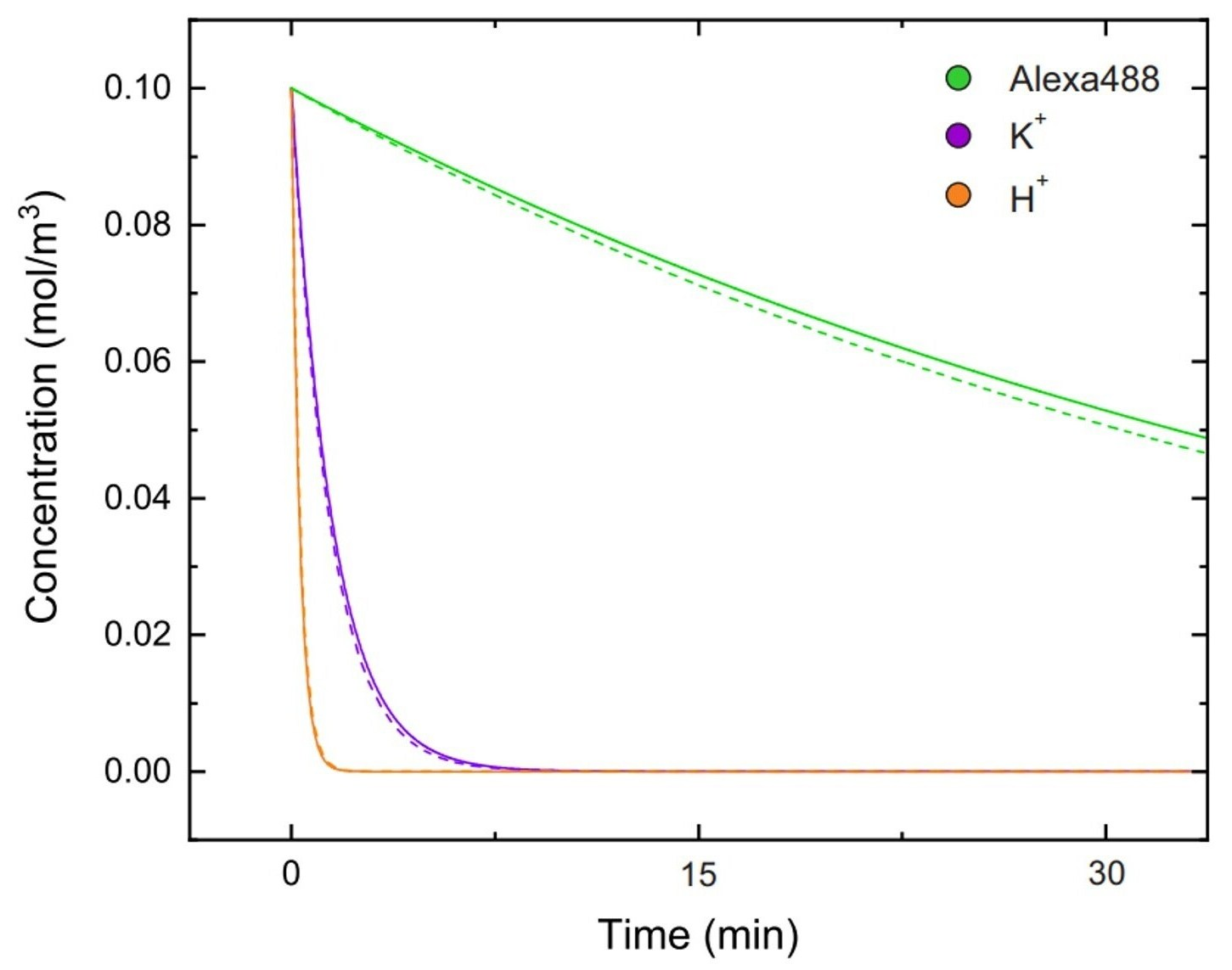}}
\caption{{\bf Modelling of diffusion through the $\alpha$-hemolysin pore.} Plot of the concentration of Alexa-488 dye (green), K$^{+}$ ions (purple) and H$^{+}$ ions (orange) inside a microwell with $5~\mu$m diameter and $600$~nm height sealed with lipid bilayer supporting a single $\alpha$-hemolysin pore for a common starting concentration of $0.1$~mol/m$^{3}$. Dashed lines represent data obtained using an analytical model based on Fick's law, solid lines represent data obtained from a finite element simulation using COMSOL Multiphysics (see supplementary text for further details). We assume diffusion coefficients of $5.1\times10^{-11}$~m$^{2}$/s for Alexa-488~\cite{WatanabeNatComm14}, $1.6\times10^{-9}$~m$^{2}$/s for K$^{+}$ ions~\cite{AllenJChemPhys99} and $7.6\times10^{-9}$~m$^{2}$/s for H$^{+}$ ions~\cite{LeeJChemPhys11}, and a pore length and diameter of $10$~nm and $1$~nm, respectively~\cite{WatanabeNatComm14}.}
\end{figure}

\subsubsection*{Modelling of diffusion rates through the $\alpha$-hemolysin pore}
Our experiment outlined in Fig.~1 of the main text aims to detect the activity of one or more $\alpha$-hemolysin pores embedded in the lipid bilayer sealing the microwell via two simultaneous means: the leakage of K$^{+}$ ions through the pore using the conductance of a PEDOT:PSS OECT at the bottom of the microwell and the leakage of Alexa-488 dye molecules through the pore using the fluorescence intensity of the microwell. These species are radically different in size, yet diffusing through a common pore, so the time-scale of these signals should be radically different. We can estimate the expected time-scale for these two signals as follows. First, we can deal with this scenario analytically using Fick's law, since the impedance of the pore should be dominant such that the concentration in the microwell is homogeneous and the concentration drops only across the pore -- The pore is of order $1$~nm diameter and $10$~nm length in a microwell of diameter $4~\mu$m and height $500$~nm. The passive transport through the pore is given by:

\begin{equation}
J = -D\frac{\partial C}{\partial z}
\end{equation}

\noindent where $J$ is the particle flux, directed to reduce the concentration gradient, and $D$ is the diffusion coefficient. The concentration gradient across the pore $\frac{\partial C}{\partial z} \simeq \frac{\Delta C}{L}$, where $\Delta C = C_{outer} -  C_{inner}$ is the concentration difference across the bilayer and $L$ is the pore length. The particle flux is also given by:

\begin{equation}
J = \frac{\partial M}{\partial t}\frac{1}{A}
\end{equation}

\noindent where $M(t)$ is the number of moles passing through the pore with a cross-sectional area $A$ in a time $t$. The concentration inside the microwell is then just $C_{inner}(t) = M(t)/V$. We can combine the two expressions for the flux $J$ to obtain:

\begin{equation}
\frac{\partial C_{inner}}{\partial t} = \frac{\partial M}{\partial t}\frac{1}{V} = -\frac{A.D}{V.L}\Delta C
\end{equation}

\noindent We can solve this differential equation to obtain:

\begin{equation}
C_{inner}(t) = \Delta C~\mathrm{exp} (-\frac{A.D}{V.L}t) + C_{outer}
\end{equation}

\noindent with $C_{inner} \rightarrow C_{outer}$ as $t \rightarrow \infty$. If there are $n > 1$ $\alpha$-hemolysin pores in the bilayer, this becomes:

\begin{equation}
C_{inner}(t) = \Delta C~\mathrm{exp} (-\frac{n.A.D}{V.L}t) + C_{outer}
\end{equation}

The second approach is to use a finite element model such as COMSOL Multiphysics to simulate the system. We used the `\textsc{passive transport of species in aqueous solutions}' model under COMSOL with a tolerance of $10^{-5}$, mostly as a check of our numbers for the analytical model. For both models, we used a pore diameter of $1$~nm~\cite{WatanabeNatComm14}, pore length of $10$~nm~\cite{WatanabeNatComm14}, Alexa-488 diffusion coefficient of $5.1\times10^{-11}$~m$^{2}$/s~\cite{WatanabeNatComm14}, K$^{+}$ diffusion coefficient of $1.6\times10^{-9}$~m$^{2}$/s~\cite{AllenJChemPhys99}, and H$^{+}$ diffusion coefficient of $5.1\times10^{-11}$~m$^{2}$/s~\cite{LeeJChemPhys11}. Note that there is some variation in all of these values across the literature; we are not aiming to be precise here only to get a feel for the relative time-scales for the two different measurements.

Figure~S7 shows the concentration inside the well $C_{inner}(t)$ for each species versus time~$t$ for a common starting concentration $C_{inner}(0)$. The half-time for the Alexa-488 dye is of order half an hour, which is consistent with the timescale for the $\alpha$-hemolysin fluorescence assays in Watanabe {\it et al.}~\cite{WatanabeNatComm14}, which is not a surprise given we use values for $D$, $L$ and $A$ that are taken from their experimental results. The crucial aspect here is the half-time for K$^{+}$ and H$^{+}$, which are drastically shorter than for the Alexa-488 dye, and of order a few minutes and less than a minute, respectively. The analytical and numerical models show good agreement.

There are three important implications that the results in Fig.~S7 have for our electrical measurements. The first is that the sampling time for the conductivity of the OECT needs to be quite short, i.e., multiple samples per minute, if is not, our result will appear as a step-change (artefact) rather than a smooth decrease (true behaviour). The second is that we need to run the experiment much longer than the electrical signal suggests in order to obtain the corresponding optical signal, i.e., for many minutes to an hour. A benefit of this is that we can clearly discriminate this from, e.g., the bilayer rupturing, in which case we would see a rapid drop in both [K$^{+}$] and [Alexa-488] (i.e., both the electrical and fluorescence signal rapidly collapse). Lastly, it highlights why we used K$^{+}$ rather than H$^{+}$ (i.e., a pH gradient) for this experiment -- it brings our electrical signal to a more sensible time-scale for accurate measurement with the apparatus we have. Detection via pH may work for membrane proteins with slower activity, e.g., ATP-ase as used in Watanabe {\it et al.}~\cite{WatanabeNatComm14}, but not so well for a membrane protein with fast activity and/or a very wide pore like $\alpha$-hemolysin. This has important implications towards future studies where a variety of other membrane proteins could be used in a device such as ours, specifically, the need for the electronic element to sense a variety of possible cations, e.g., K$^{+}$, Na$^{+}$, Ca$^{2+}$, and not just H$^{+}$. This provides a strong incentive to design such devices using organic electrochemical transistors, where all cations and not exclusively H$^{+}$ affect conductivity, rather than semiconductor ISFETs, which sense via protonation of an oxide, and thus only really sense H$^{+}$ well.

\subsubsection*{Additional Data from Simultaneous Electrical/Optical Detection Assays}
Supplementary Figures~S8-S12 present additional data from $\alpha$-hemolysin assays to demonstrate simultaneous optical and electrical detection of membrane protein activity. Supplementary Figure~S8 provides some additional data to Figure~5a of the main text regarding the various possible behaviours of microwells during the assay. Supplementary Figure~S9 provides conductance data for OECTs with two different pre-assay hydration protocols to demonstrate how hydration affects the tail-behaviour of the conductance traces in particular. Supplementary Figure~S10 shows control data for our experiment comparing three devices from Fig.~5 of the main text with three devices from a separate assay where the $\alpha$-hemolysin concentration of the $\alpha$-hemolysin solution is zero to demonstrate a baseline for change in fluorescence and conductance in a bilayer-sealed device microwell. Supplementary Figures~S11 and S12 provide additional intensity and conductance data from device microwells performed on separate coverslips and in separate assays to demonstrate the reproducibility of what we observe in Fig.~5 of the main text.

\subsubsection*{Additional information regarding experimental methods.}
The experimental methods used are comprehensively discussed in the Materials and Methods section, which refers to some supplementary figures for additional detail. These supplementary figures are presented below.

Figure~S13 shows an overlay of the three photomasks used in the fabrication of our devices. These are provided to help the reader understand the layout and scale of the various features of the device. The alignment markers are carefully designed to ensure good alignment since this is crucial to successful device operation, mostly because the CYTOP layer serves also as the electrical isolation between the metal interconnects and the fluid in the microwells/flow-cell. If alignment is poor, parts of the electrodes are exposed and the current leakage via any liquid makes the measurement impossible. The `UP' marker is well worth its `real estate' cost, significantly improving both ease of fabrication/handling, and as a result, device yield. The common source architecture is implemented via a `C' pad at each end of the chip for two reasons. The first is redundancy, since it is an interconnect used for every device measurement, unlike the drain line, which is used on an individual basis, a cut in the source line will render the entire coverslip unusable. Having two pads means that if one is broken, the other can be used. The second is that it gives us a `straight through' path that enables us to check/monitor the metallisation resistance from batch to batch.

Figure~S14 shows the mechanism for electrical contact with the device coverslip. Bond wires are completely impractical for these structures, forcing us to direct mechanical contact via spring pins. The quantity of spring pins involved means that a significant `spring force' is exerted on the coverslip if the pins are not very well aligned in height. We achieve this using a pair of small PCBs that trap a `raised nodule' on the spring pin housing from above and below. The housings are mounted in this PCB-pair and screwed into place before the upper PCB is positioned and screwed down, and then the housings are soldered to the upper PCB. This brings the equilibrium heights of the spring pins to a common level and ensures they engage well at the contact pads. Before we took this approach we had endless problems getting $100\%$ contact pad connection and broke $>75\%$ of our coverslips during or even before measurement. After we took this approach, our breakage loss is close to zero and we mostly get $100\%$ contact pin connection assuming the mount is screwed down flat and level (we use a set of machined `height setters' to assist with this). We find the serated tip Harwin pins work best. They need frequent cleaning, irrespective, to prevent crud on the tips from interfering with electrical contact.

Figure~S15 shows a photograph of the measurement circuit with a device under measurement. The custom-built multiplexer features a set of four MAX306 high-performance 16-channel CMOS analog multiplexer chips (Analog Devices). The multiplexers are controlled using the GPIO pins of a Raspberry Pi (mounted on top of the multiplexer box), which receives its instructions from the control PC via TCP/IP (ethernet). The MAX306 has a through-resistance that is dependent on the supply voltage. We found performance poor if we tried to run these with a unipolar $+5$~V supply (i.e., off the Raspberry Pi itself), and much better with a bipolar $+/-15$~V supply. For convenience and lower noise, we ran the MAX306 chips off a battery supply ($4\times9$~V batteries), with a small circuit to step the battery-set ($+/-18$~V nominal) down to $+/-15$~V rails for the MAX306s.

Figure~S16 shows the design and assembly of our flow-cell. The two holes are drilled with a diamond-tipped drill, which is a highly labour intensive process that needs to be done with significant care. This is mostly because drilling too fast either burns the hole and tip or results in cracking of the block. Given the opportunity again, laser cutting the block and holes, rather than drilling holes into a prefabricated block would be the better option (at the very least, ensure you have a good audiobook ready prior to embarking on the many mind-numbing hours required to drill those holes).

Figure~S17 shows a photograph of our device and flow-cell assembly under measurement conditions. Figure~S18 shows a schematic of the microfluidics system for the experiment. Figure~S19 shows a schematic of the microscopy optics for the experiment. Figure~S20 shows a typical $1214\times1214~\mu$m field-of-view taken with our microscopy system during a typical $\alpha$-hemolysin assay, which is sufficient to monitor all $52$ OECT device wells along with a large number of non-device wells without needing to adjust stage positioning.

\begin{figure}
\centering
{\includegraphics[width = 0.75\textwidth]{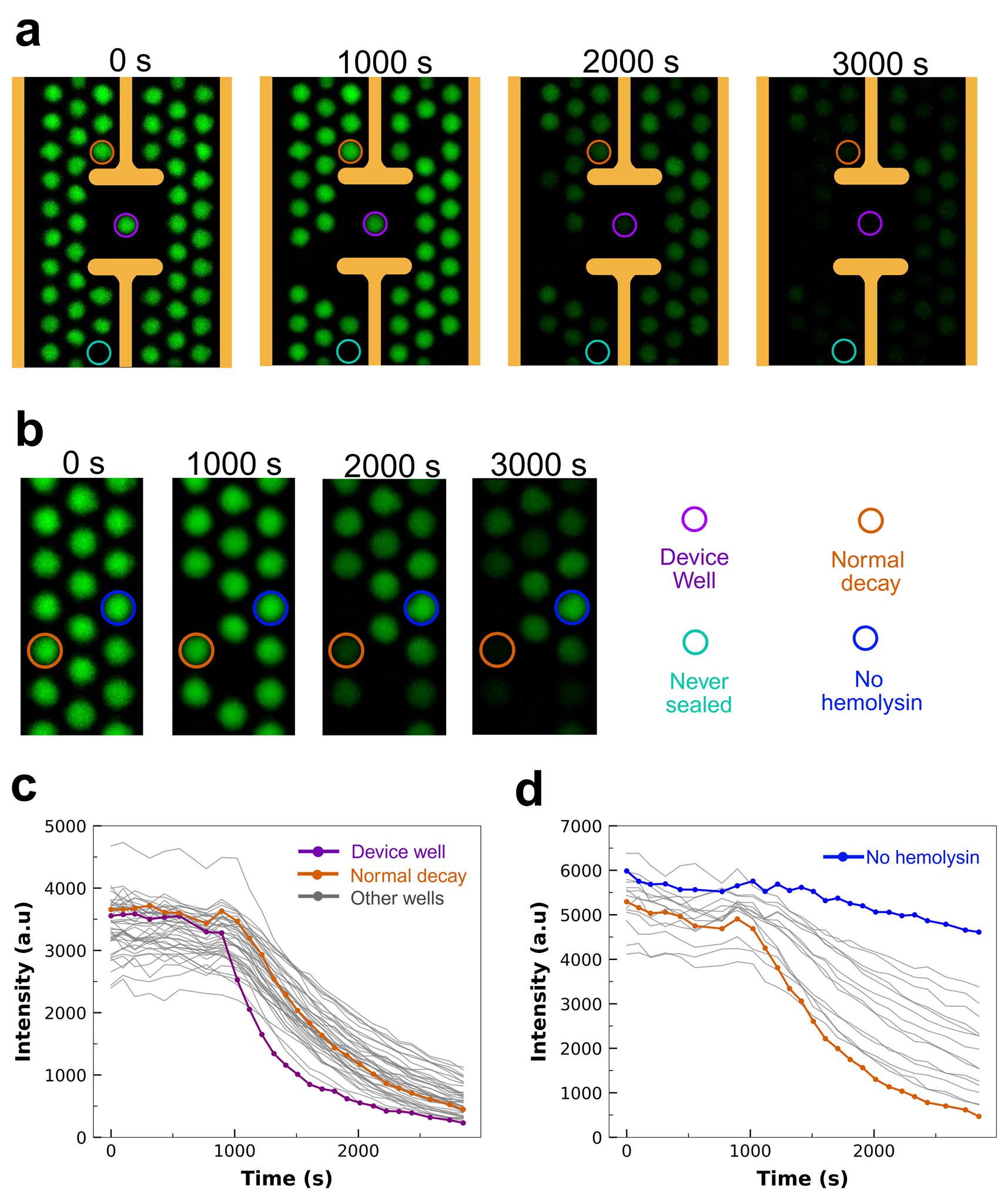}}
\caption{{\bf Additional fluorescence microscopy data showing response of a microwell without $\alpha$-hemolysin pore.} ({\bf a}) Data from a separate device assay showing the device microwell (purple circle), and non-device microwells that sealed and show normal decay due to embedded $\alpha$-hemolysin (brown circle), and that never sealed (cyan circle). ({\bf b}) Non-device wells showing normal decay (brown circle) and response of a microwell that sealed but had no embedded $\alpha$-hemolysin (blue circle). ({\bf c}) Plots of fluorescence intensity versus time for the microwells in ({\bf a}) with the circled traces in ({\bf a}) highlighted and traces for any microwell that ruptured during the assay removed for clarity. ({\bf d}) Plots of fluorescence intensity vs time for the microwells in ({\bf b}) with the circled traces in ({\bf b}) highlighted and traces for any microwell that ruptured during the assay removed for clarity.}
\end{figure}

\begin{figure}
\centering
{\includegraphics[width = 1.0\textwidth]{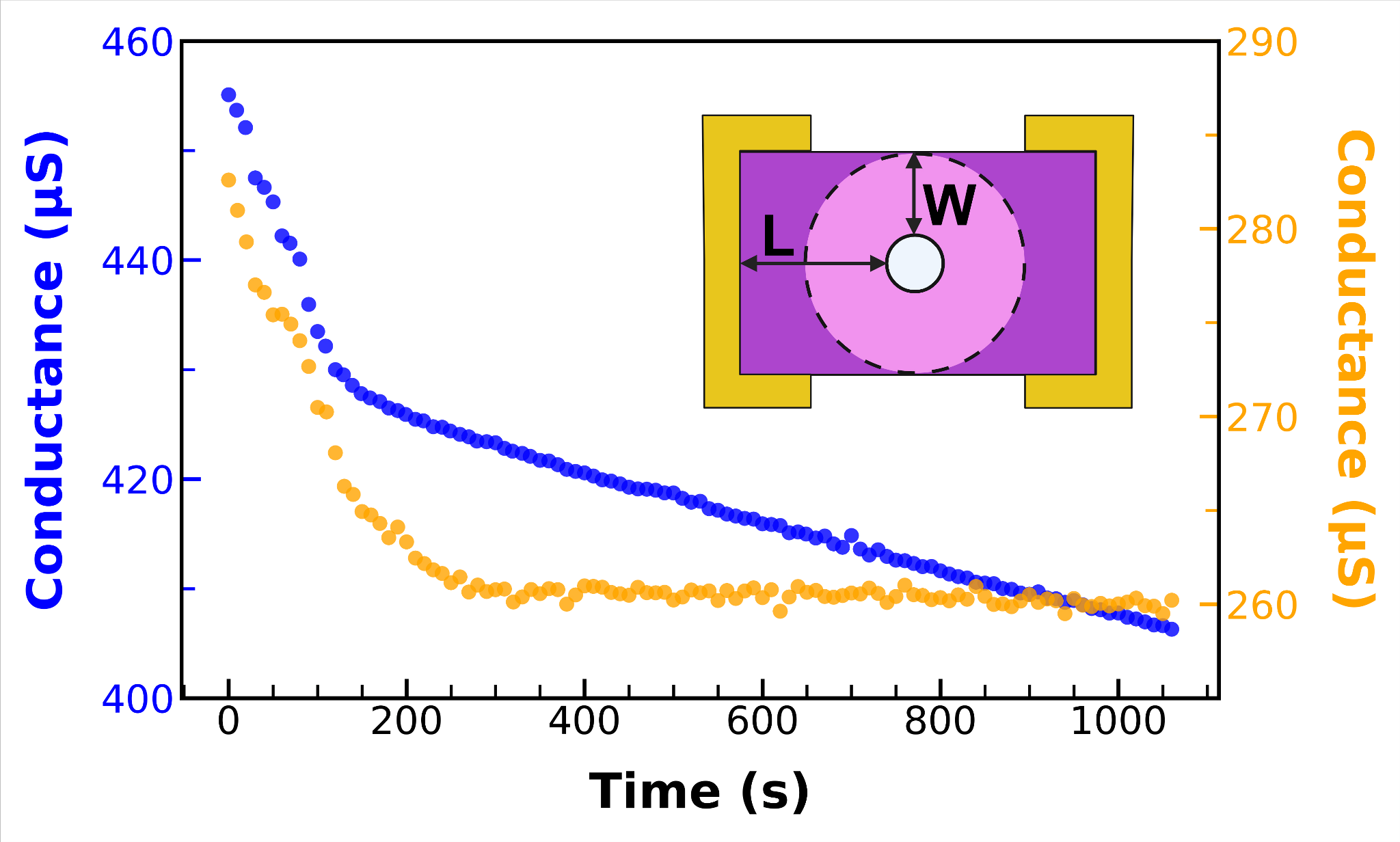}}
\caption{{\bf Effect of device hydration prior to the assay on the device response in assays.} Plots of the measured OECT conductance vs time for OECTs subject to pre-hydration for $3$~hours (blue/left axis) and $5$~hours (orange/right axis). Inset shows a schematic of an OECT showing the ion-concentration-change diffusion front (dashed circle/pink), which reaches the side of the OECT microwell before it reaches the source and drain contacts. The ratio of the `side-segment' width $W$ to the half-channel length $L$ is a key design parameter in determining the response of our OECTs to a change in K$^{+}$ ion concentration in the microwell.}
\end{figure}

\begin{figure}
\centering
{\includegraphics[width = 1.0\textwidth]{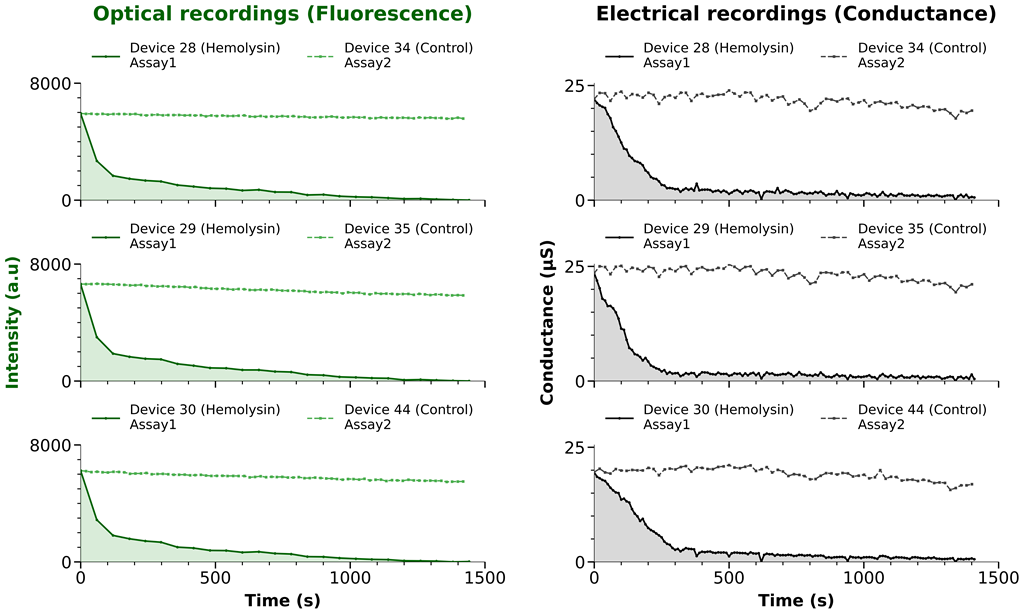}}
\caption{{\bf Control data for an assay with [$\alpha$-hemolysin] = 0.} Plots of fluorescence intensity vs time (left column/green) and OECT conductance vs time (right column/green) for three devices from Figure~5 of the main text (labelled `hemolysin') and three devices from a separate assay run where [$\alpha$-hemolysin] = 0 in the assay solution (labelled `control'). Data shown indicates the baseline for changes in fluorescence and conductance in lipid-sealed microwells in Figure~5 of the main text.}
\end{figure}

\begin{figure}
\centering
{\includegraphics[width = 0.8\textwidth]{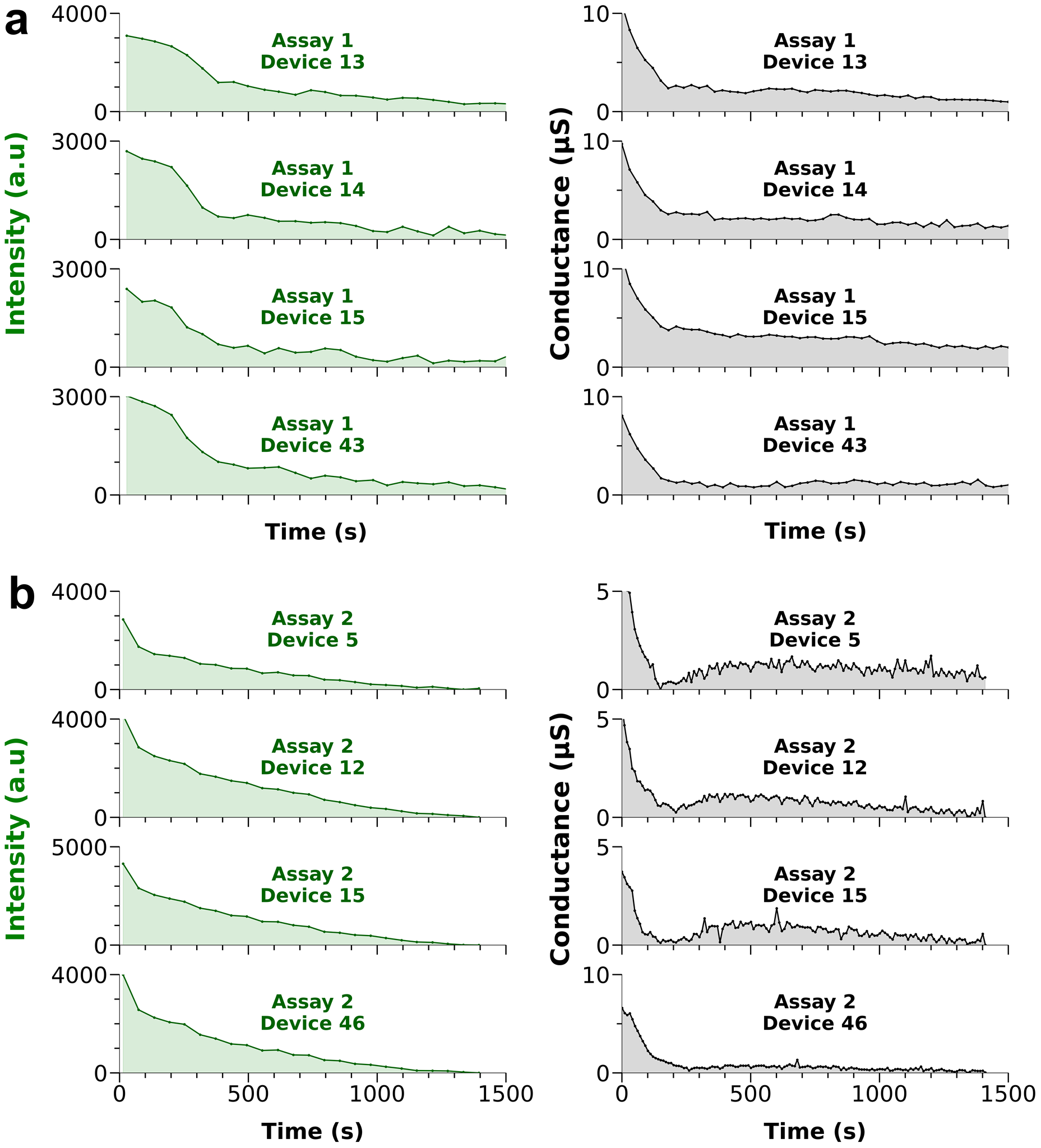}}
\caption{{\bf Additional data from two additional assays on separate coverslips.} ({\bf a,b}) Plots of fluorescence intensity vs time (left column/green) and OECT conductance vs time (right column/green) for four OECT devices on two separate coverslips and two separate assay runs. The intensity/conductance data has an offset of ({\bf a}) Device~13: $2234$~a.u./$205~\mu$S, Device~14: $1864$~a.u./$206~\mu$S, Device~15: $1614$~a.u./$206~\mu$S, Device~43: $1881$~a.u./$216~\mu$S, and ({\bf b}) Device~5: $3391$~a.u./$239~\mu$S, Device~12: $4032$~a.u./$250~\mu$S, Device~15: $4102$~a.u./$275~\mu$S and Device~46: $3383$~a.u./$242~\mu$S subtracted for clarity. Data shown here is intended to demonstrate reproducibility of our main result in Figure~5 of the main text.}
\end{figure}

\begin{figure}
\centering
{\includegraphics[width = 0.8\textwidth]{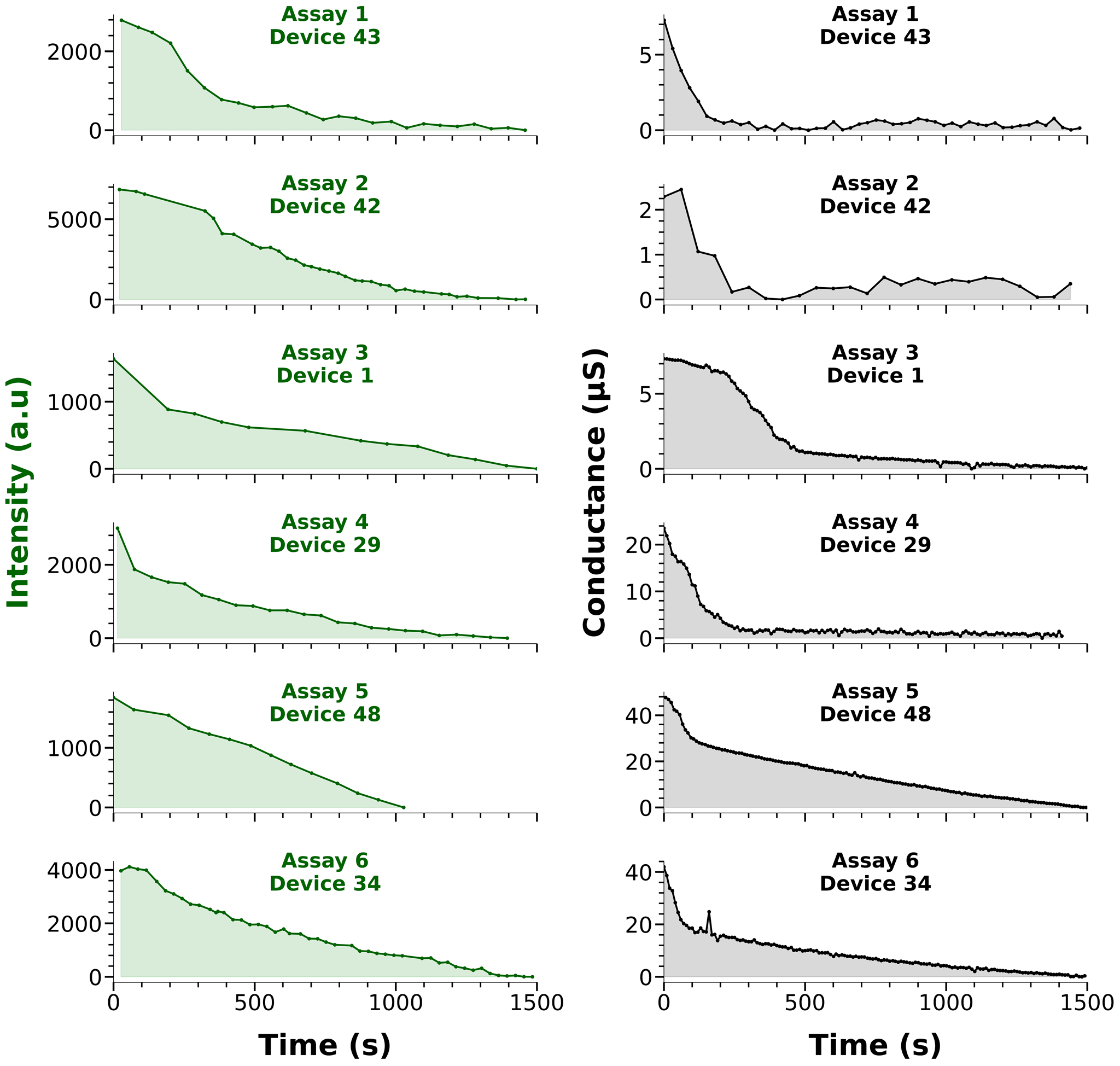}}
\caption{{\bf Additional data from six additional assays on separate coverslips.} Plots of fluorescence intensity vs time (left column/green) and OECT conductance vs time (right column/green) for a single selected OECT device on six separate coverslips and six separate assay runs. The intensity/conductance data has an offset of Device~43: $2116$~a.u./$216~\mu$S, Device~42: $924$~a.u./$120~\mu$S, Device~1: $1411$~a.u./$134~\mu$S, Device~29: $3382$~a.u./$259~\mu$S, Device~48: $807$~a.u./$400~\mu$S, Device~34: $5259$~a.u./$353~\mu$S subtracted for clarity. Data shown here is intended to present some of the variability in outcomes we see across the series of assay runs we performed for this work.}
\end{figure}

\begin{figure}
\centering
{\includegraphics[width = 1.0\textwidth]{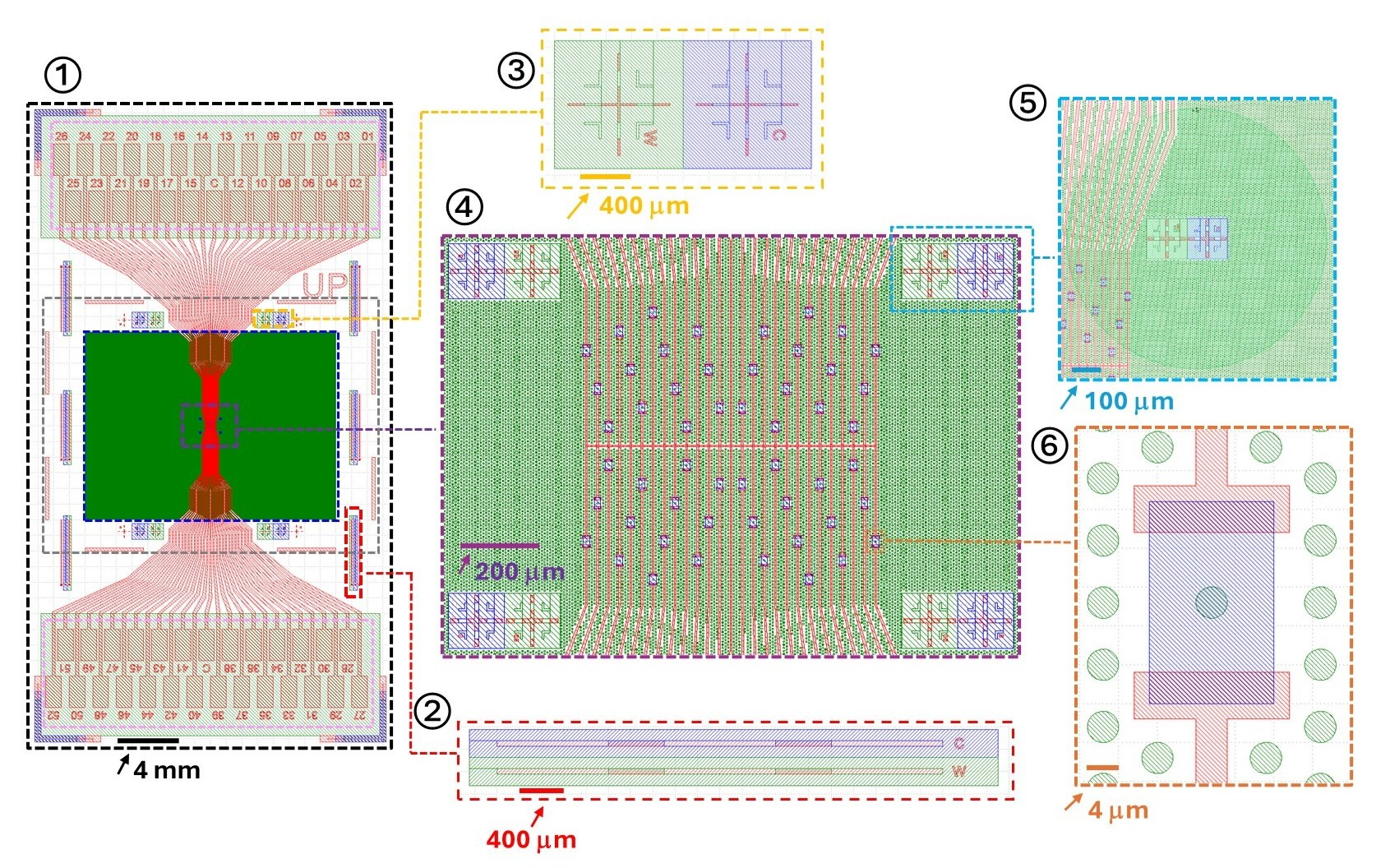}}
\caption{{\bf Composite photomask overlay to illustrate device design.} An overlay is presented of the three photomasks used to make our devices. The first mask (red) is used to define the Al/Au layer, the second mask (blue) is used to define the PEDOT:PSS OECT channels, and the third mask (green) is used to define the hydrophobic CYTOP layer. The full mask is shown in \ding{172}. Working from the outside inwards, the black dashed lines demarcate the $24\times40$~mm coverslip itself (outer), the region covered by the $20\times15$~mm glass block forming the flow-cell, the region containing the hexagonally close packed array of $4~\mu$m diameter wells, and lastly the region shown in \ding{175} (inner). The remaining panels are magnifications of subcomponents of the mask including the angle-alignment markers in \ding{173}, the coarse position-alignment markers in \ding{174}, the active device region including the fine position-alignment markers in \ding{175}, the fine position-alignment markers themselves in \ding{176} and finally one of the 52 devices in \ding{177}. The lengths of the various scale-bars are indicated on the diagram.}
\end{figure}

\begin{figure}
\centering
{\includegraphics[width = 1.0\textwidth]{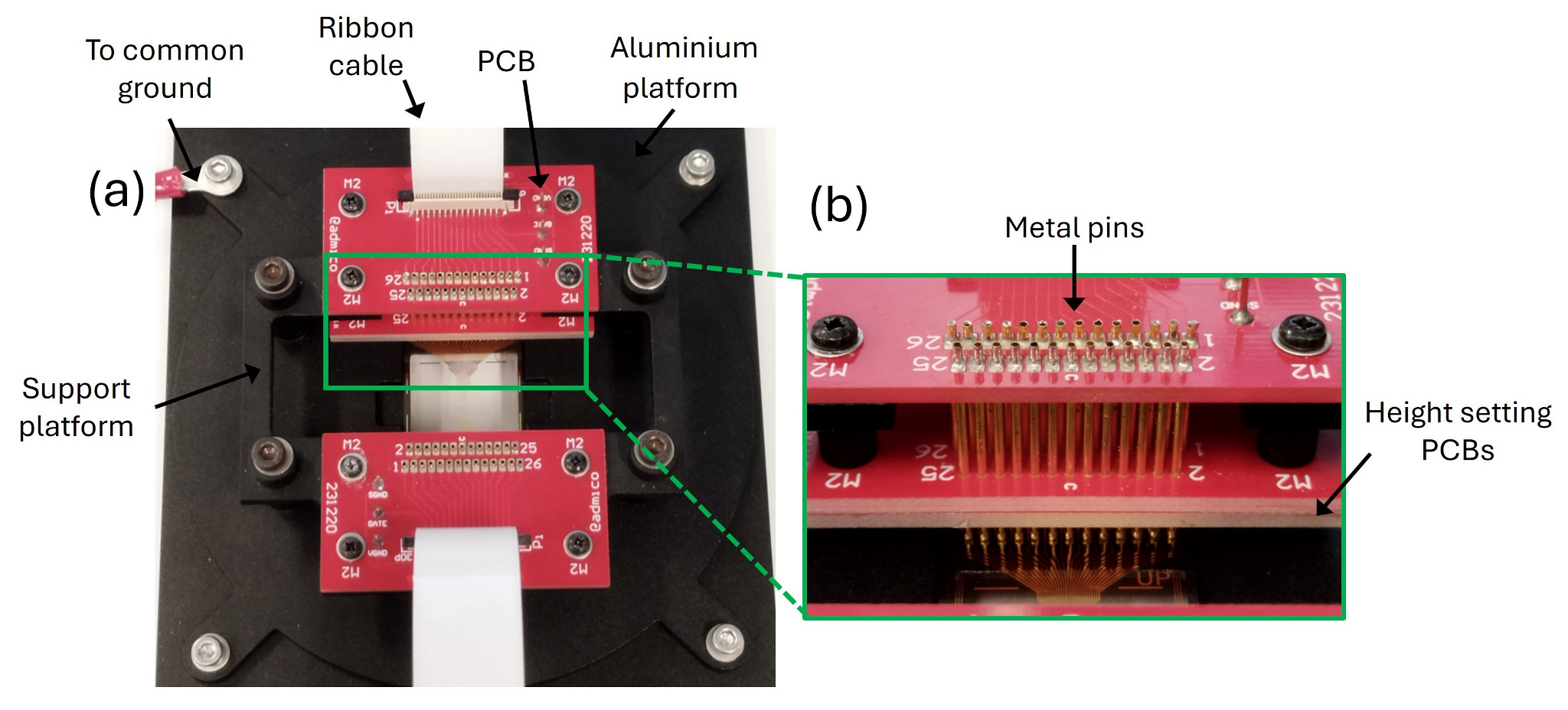}}
\caption{{\bf Electrical contact to the device coverslip.} ({\bf a}) Photograph of the apparatus used for making electrical contact to the device coverslip on the microscope. A custom CNC-machined aluminium bracket is used to mount a custom printed-circuit board that connects a 30-wire flat flexible cable (Samtec FJH-30 series) to a set of 27 spring pins (Harwin S13-503/P13-4023), which are held with their equilibrium position as close to a flat plane as possible using a pair of `height setting' printed circuit boards, shown in the zoom image in ({\bf b}). The remaining 3 lines on the flat flexible cable are used for gate and ground lines. There are 27 pins on each end making a total of 54 connections to the coverslip.}
\end{figure}

\begin{figure}
\centering
{\includegraphics[width = 0.8\textwidth]{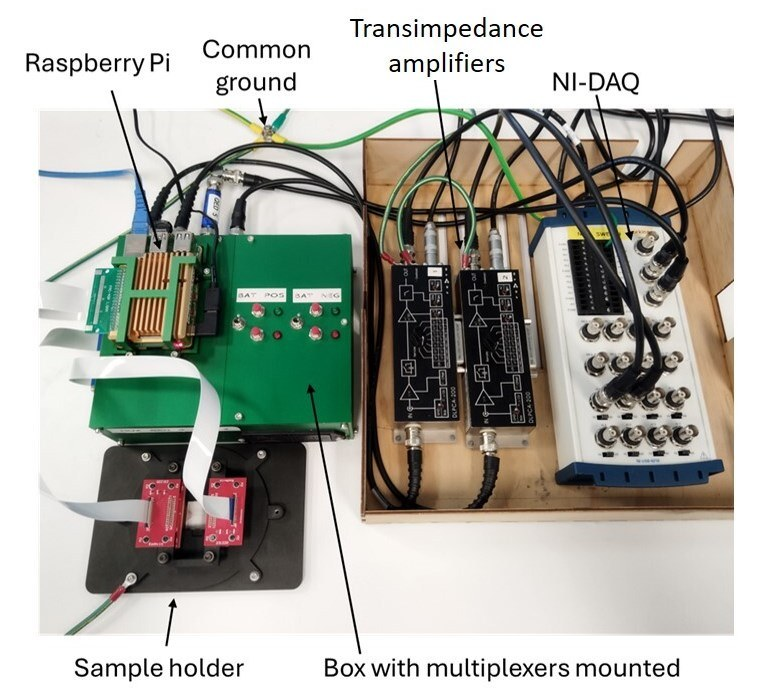}}
\caption{{\bf Photograph of measurement circuit and device under measurement.} The two 30-line flat flexible cables that connect to the 54 contact pads of the device coverslip (see Fig.~S13) connect to a custom-built multiplexer system (green box). The multiplexer chips operate in pairs to sequentially connect pins from each of the two ends of the device coverslip to one of the two transimpedance amplifiers (Femto DLPCA-200), which convert each drain current $I_{d}$ to a voltage that is read at two analog inputs of a National Instruments USB-6216 data acquisition device. The USB-6216 is also used for its analog outputs, which supply the source-drain bias $V_{sd}$ and gate voltage $V_{g}$ that pass via the multiplexer box and 30-pin flat flexible cables. The source-drain voltage $V_{sd}$ is applied to the central `C' pad at both ends of the device coverslip. The gate voltage $V_{g}$ terminates at a pin on the sample holder printed circuit board for connection to the gate electrode (see Fig.~S14). The various instruments and other components are carefully grounded to minimise electrical noise. The device is often enclosed in a custom-machined enclosure (not shown) during measurement, which acts as a Faraday shield to further reduce noise.}
\end{figure}

\begin{figure}
\centering
{\includegraphics[width = 1.0\textwidth]{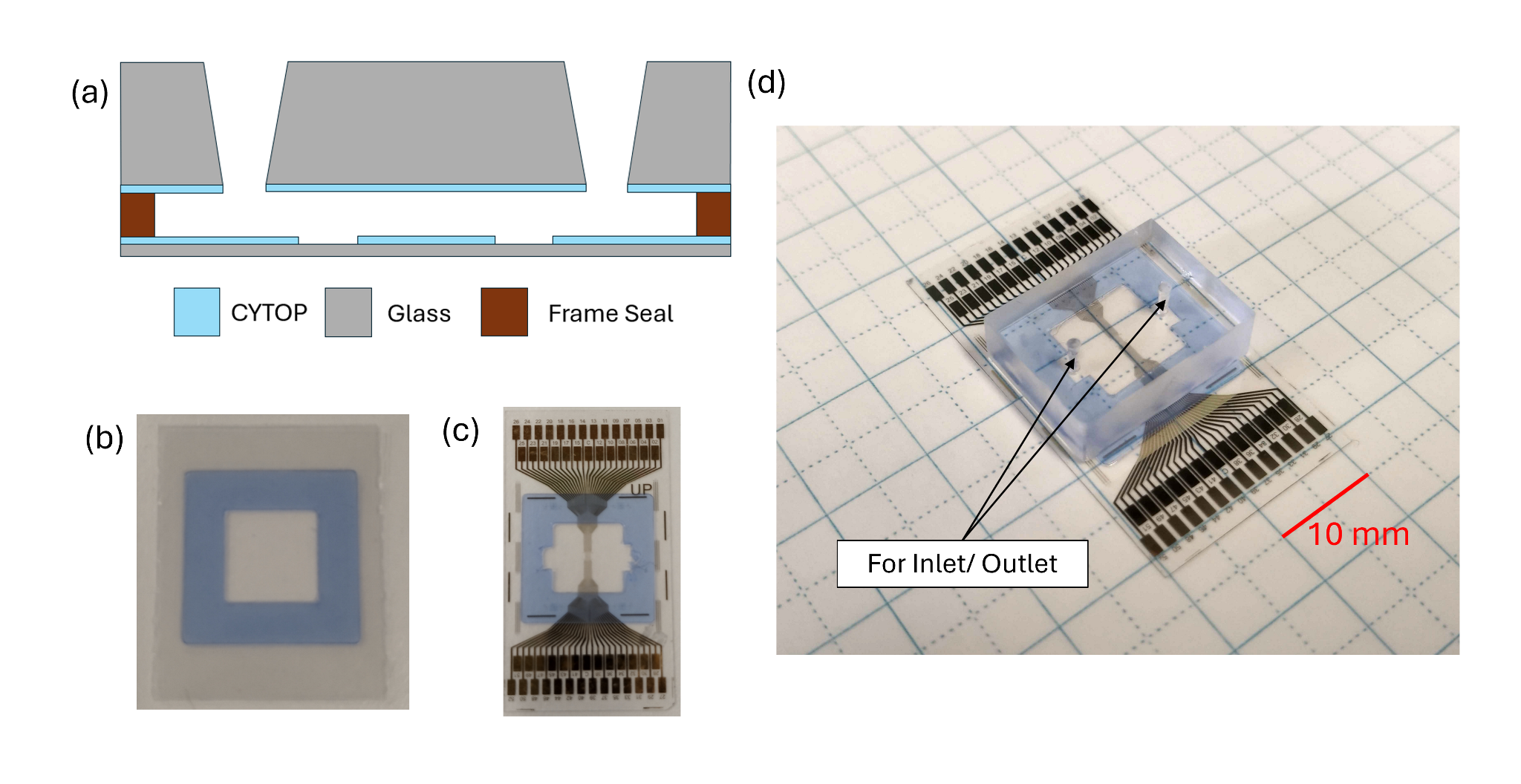}}
\caption{{\bf Flow-cell design and assembly.} ({\bf a}) Side-view schematic of our flow-cell structure (not to scale). A CYTOP-coated $20\times15\times5$~mm BK7 glass block (Eksma Optics) with two holes drilled in it is adhered to our completed device coverslip using a `frame-seal' (Biorad SLF0201), which is shown in ({\bf b}). ({\bf c}) We make two small rectangular cuts into the sides of the frame-seal and adhere it to the glass coverslip. ({\bf d}) We then adhere the CYTOP-coated glass block over the top to form a closed flow-cell with two access ports via the glass block.}
\end{figure}

\begin{figure}
\centering
{\includegraphics[width = 0.8\textwidth]{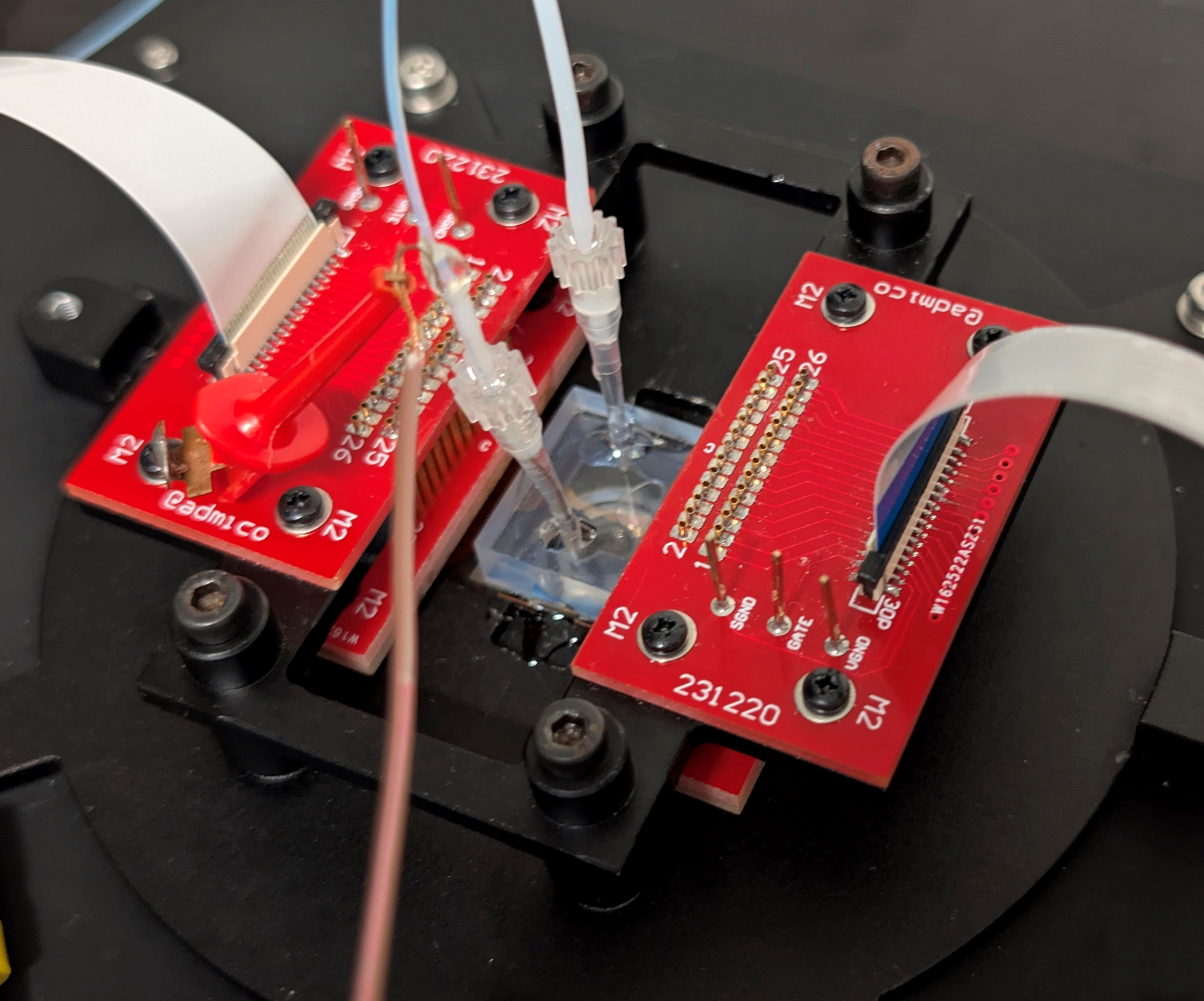}}
\caption{{\bf Photograph of device coverslip under measurement conditions.} The device coverslip with the completed flow-cell (see Fig.~S16d) is mounted on the microscope and held in place using the spring pin apparatus for electrical contact (see Fig.~S14a). The fluidic connections are made to the flow-cell by gluing the fluidic lines into the pipette tips and then the pipette tips into the ports in the glass block using UV-curable epoxy (Norland NOA68). To manage the electrochemical potential of the outer solution in the flow-cell, we thread a $230~\mu$m diameter Ag wire, the end of which has been chloridised, down via the gap between the fluidic hose and the pipette tip and into the end of the flow-cell, as shown in the photograph.}
\end{figure}

\begin{figure}
\centering
{\includegraphics[width = 1.0\textwidth]{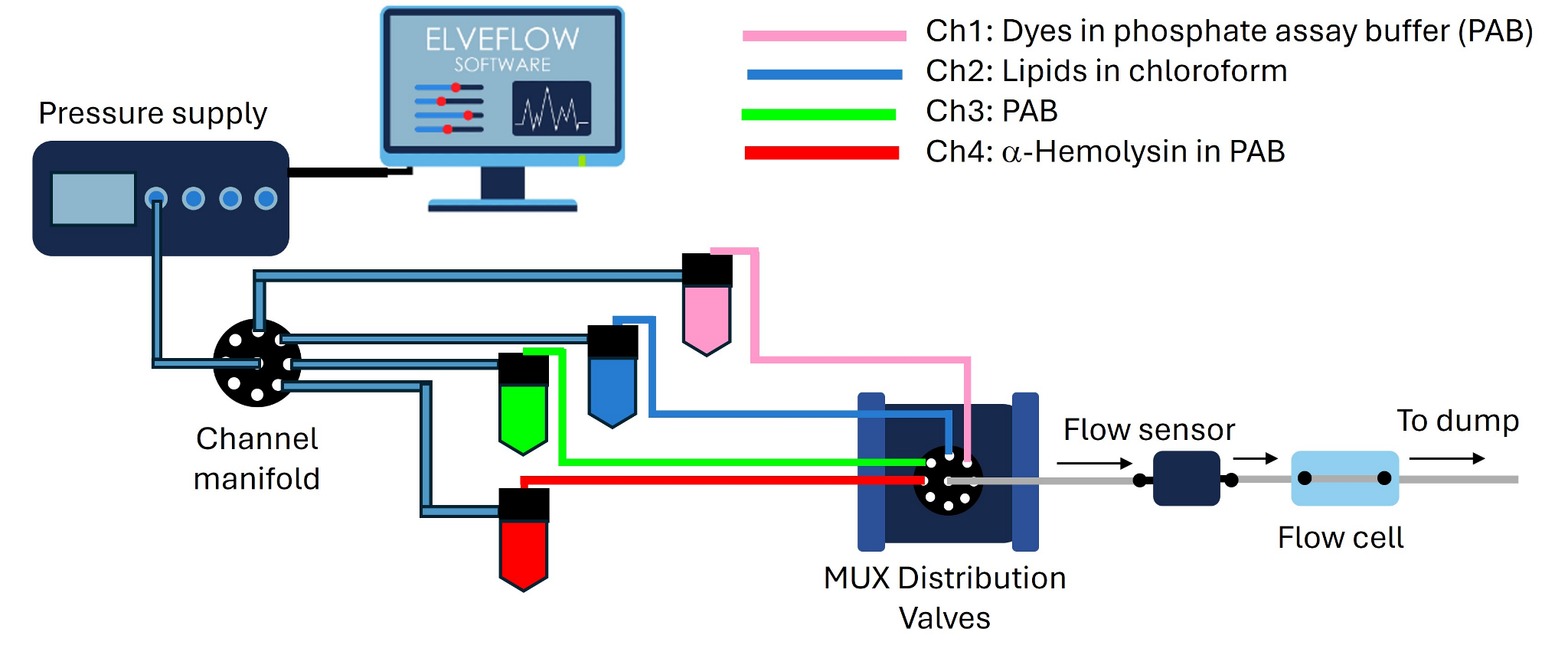}}
\caption{{\bf Microfluidics system for the experiment.} We use an Elveflow pressure/flow controller system (OB1 MK4) with multiplexed distribution lines (12/1 rotary bidirectional microfluidic valve) to feed the flow-cell from four different reservoirs/channels. Channel 1 contains the inner solution (phosphate buffer with Alexa-488 dye and $50$~mM KCl), Channel 2 contains the lipid in chloroform solution, Channel 3 contains the outer solution (dye-free phosphate buffer with $100$~mM KCl), Channel 4 is outer solution with $5~\mu$g/mL $\alpha$-hemolysin added ($150$~nM).}
\end{figure}

\begin{figure}
\centering
{\includegraphics[width = 1.0\textwidth]{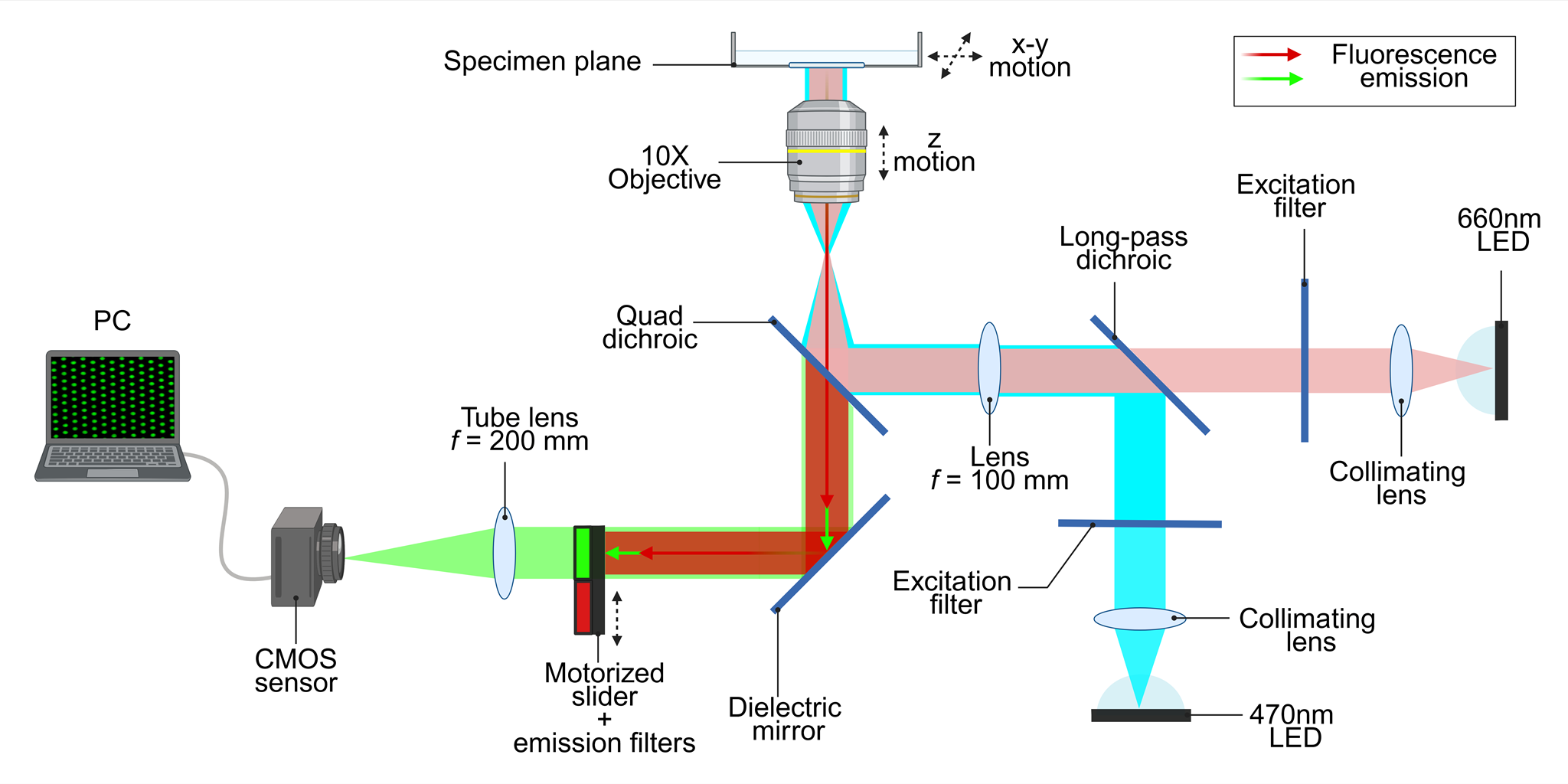}}
\caption{{\bf Microscopy optics for the experiment.} We use a custom-built fluorescence microscope for our experiment featuring two LED sources at $470$~nm (Thorlabs M470L5) and $660$~nm (Thorlabs M660L4), each with a dichroic filter in front at $495$~nm  (Chroma ET495) and $645$~nm (Chroma ET645), respectively. These two beams can be combined using a dichroic beam combiner (Nikon DM500) prior to collimation using a $100$~mm focal length collimating lens. The excitation then passes to the device via a $10\times$ objective lens (Nikon Plan-NEOFLUAR) giving a nearly collimated beam at the objective lens output. The emission signal returns via the objective and a quad dichroic mirror (Semrock BrightLine 405/488/561/635~nm) placed between the objective lens and collimating lens. The beam then passes through an $525$~nm or $690$~nm dichroic filter (Chroma ET525/ET690) mounted on a motorized slider followed by a $200$~mm focal length tube lens. The tube lens image is projected onto the CMOS sensor (ZWO ASI2600MM Pro with $6248\times4176$~pixel CMOS sensor). The device coverslip is mounted in a custom-built microscope sub-stage mounted on a Mad City Labs two-axis servo-controlled stage used to translate the sample along the $x$-$y$ plane using a controller (MCL Micro-Drive). For focus control, the objective is mounted on an L-shaped bracket connected to a vertical translation stage driven by a DC servo motor (Thorlabs MT1/Z825B/KDC101) to enable computer-controlled fine-focus adjustment during the assays.}
\end{figure}

\begin{figure}
\centering
{\includegraphics[width = 1.0\textwidth]{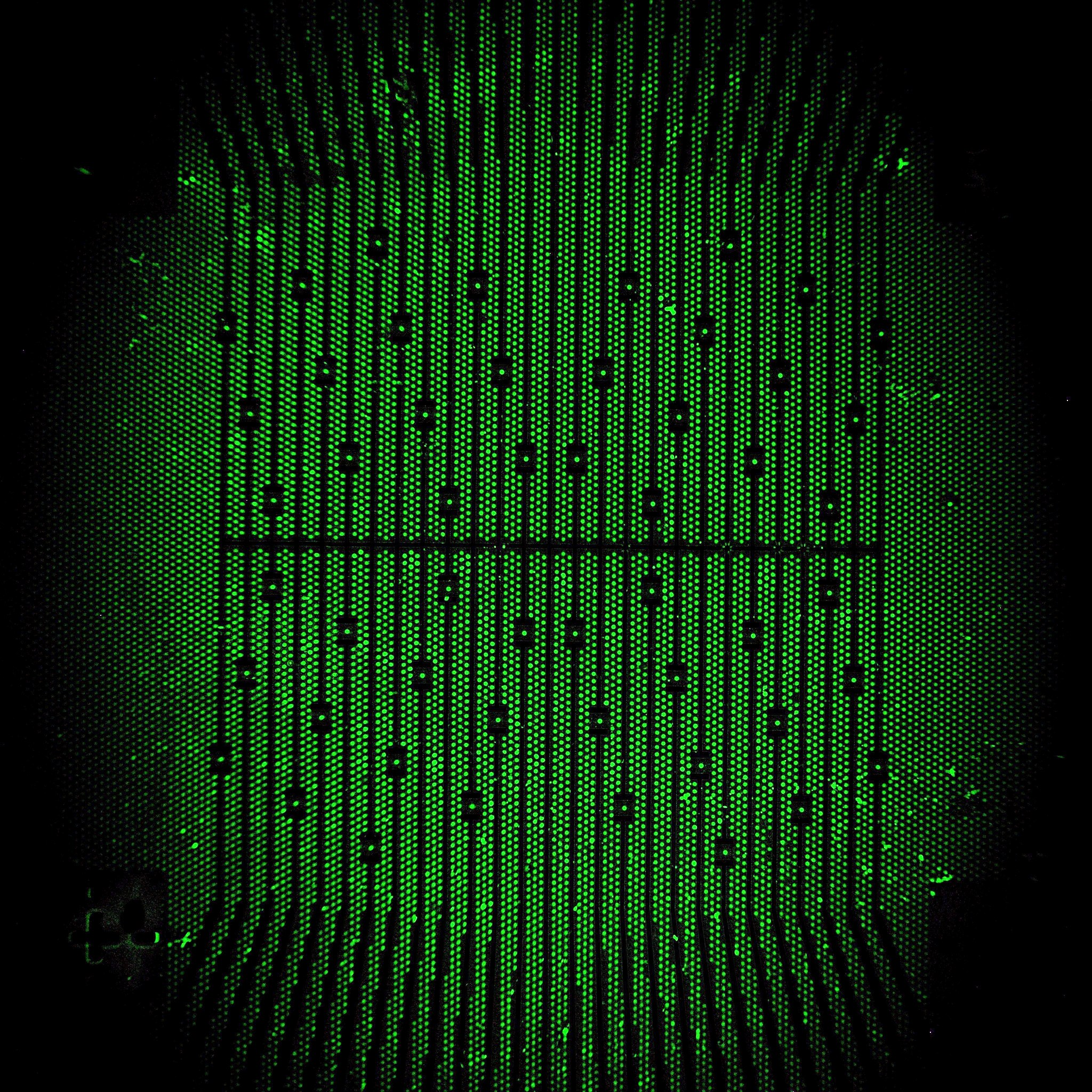}}
\caption{{\bf Typical $\alpha$-hemolysin assay fluorescence image.} The image shows the full $1214\times1214~\mu$m field-of-view of a device coverslip taken during an $\alpha$-hemolysin assay presented `as obtained' with the greyscale converted to greenscale for clarity.}
\end{figure}

\end{document}